\documentclass[journal]{IEEEtran}
\usepackage{blindtext}
\usepackage{graphicx}
\ifCLASSINFOpdf
\else
\fi
\usepackage[cmex10]{amsmath}

\usepackage{subfig}

\DeclareMathSizes{10}{10}{6}{6}
\usepackage{amsmath,bm}
\usepackage{csquotes}
\usepackage{accents}
\newcommand{\ubar}[1]{\underaccent{\bar}{#1}}
\usepackage{amsthm}
\newtheorem{theorem}{Theorem}

\usepackage{graphicx}  
\newcommand{\blue}[1]{\textcolor{black}{#1}}
\newcommand{\red}[1]{\textcolor{red}{#1}}
\usepackage{stfloats}

\usepackage{gensymb}
\usepackage{amsmath}
\usepackage{algorithm}
\usepackage{algpseudocode}
\usepackage{latexsym}
\usepackage{mathrsfs}
\usepackage{tabularx}
\usepackage[symbol]{footmisc}
\usepackage{footnote}
\usepackage{blkarray}
\usepackage{multirow}
\usepackage{tabstackengine}
\usepackage{tikz,tikz-3dplot}
\usetikzlibrary{arrows.meta}
\usetikzlibrary{%
    decorations.pathreplacing,%
    decorations.pathmorphing%
}
\usepackage{verbatim}
\usepackage[american,cuteinductors,smartlabels]{circuitikz}
\usetikzlibrary{math}
\usepackage{amsmath}
\tikzset{%
    body/.style={inner sep=0pt,outer sep=0pt,shape=rectangle,draw,thick,pattern=north east lines wide},
    dimen/.style={<->,>=latex,thin,every rectangle node/.style={fill=white,midway,font=\sffamily}},
    symmetry/.style={dashed,thin},
}
\usepackage{hyperref}
\hypersetup{
    colorlinks,
    linkcolor={red!50!black},
    citecolor={blue!50!black},
    urlcolor={blue!80!black}
}

\begin{document}

\title{General Theory of Coupled Characteristic Mode: An Eigen Subspace Approach}

\author{Rakesh~Sinha\textsuperscript{\href{https://orcid.org/0000-0003-0592-8505}{\includegraphics[scale=0.1]{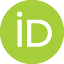}}},~\IEEEmembership{Member,~IEEE} and Sandip~Ghosal\textsuperscript{\href{https://orcid.org/0000-0003-1222-8013}{\includegraphics[scale=0.1]{fig/orcid.png}}},~\IEEEmembership{Member,~IEEE} \vspace*{-2em}
        % <-this % stops a space
\thanks{Manuscript received Mar. 26, 2025.  }
\thanks{Rakesh Sinha is with the Department of Electrical Engineering, National Institute of Technology, Rourkela-769008, Odisha, India (e-mail: r.sinha30@gmail.com).}
\thanks{Sandip Ghosal is with Department of Electronics and Communication Engineering, National Institute of Technology, Rourkela-769008, Odisha, India  }
\thanks{Color versions of one or more of the figures in this paper are available online
at http://ieeexplore.ieee.org.}
\thanks{\red{This work has been submitted to the IEEE for possible publication.
Copyright may be transferred without notice, after which this version
may no longer be accessible.}}}


\markboth{IEEE Transactions on Antenna and Propagation}{SINHA and GHOSAL: General Theory of Coupled Characteristic Mode: An Eigen Subspace Approach }



\maketitle

\begin{abstract}
%\boldmath
In this work, the problem of characteristic mode
analysis using eigendecomposition of the method of moments
impedance matrix has been simplified using the eigen-subspace
approach. The idea behind the eigen-subspace arises from the
physical properties of antenna or scatterers, where only a few
eigenmodes are enough to characterize the antenna or scatterer.
Therefore, entire space eigenanalysis is a waste of computational
resources, and eigen-subspace analysis with few modes is good
enough to characterize antennas and scatterers. It has been
assumed that there is an eigen-subspace (or hyperplane) of
coupled characteristic mode, which coincides with the eigen-
hyperplane of uncoupled characteristic mode. We can say the coupled characteristic modes are
linear combinations of isolated modes based on this assumption. The linear combination is mapped via \textit{modal coupling matrix}. Using the \textit{modal coupling matrix}, we can explain the behavior of arbitrarily shaped antennas and scatterers.  A computationally efficient
method is developed to compute coupled characteristic modes of
two mutually coupled scatterers or antennas using the eigen-
subspace. The method is summarized as a theorem of two-body coupled characteristic mode. The theorem of two-body
coupled characteristic mode has been extended to the N-body coupled
characteristic mode. Two algorithms have been developed for
the two-body multimode coupled characteristic mode and N-body
multimode coupled characteristic mode. Two numerical examples
are provided to validate the proposed concepts.
\end{abstract}

% \color{blue}
\section*{Nomenclature}
\addcontentsline{toc}{section}{Nomenclature}
\begin{IEEEdescription}[\IEEEusemathlabelsep\IEEEsetlabelwidth{$\textbf{R}_{N\times N}$ and $\textbf{X}_{N\times N}$}]
 \item [${[\textbf{Z}]_{N\times N}}$] Method of moments Z-matrix with N basis function.
 \item [${[\textbf{R}]_{N\times N}}$ and ${[\textbf{X}]_{N\times N}}$] Re. and Im. part of MoM self impedance matrix $[\textbf{Z}]=[\textbf{R}]+j[\textbf{X}]$.
\item [${[\text{V}]_{N\times 1}}$ and ${[\text{I}]_{N\times 1}}$] Excitation vector and surface current coefficient vector in MoM formulation $[\textbf{Z}][\text{I}]=[\text{V}]$.
\item [${[\textbf{I}]_{N\times N}}$]  $=[I_1,I_2,..,I_N]$: Set of N eigenvectors covers entire eigenspace of characteristic mode equation ($I_k : (\textbf{X} -\lambda_k \textbf{R})I_k = 0$) with $[I_k]$ is the $k^{th}$ characteristic mode.
\item [${[\bm{\lambda}]_{N\times N}}$]  $=diag(\lambda_1,\lambda_2,..,\lambda_N)$: All the eigenvalues of characteristic mode equation with $\lambda_k$ is the $k^{th}$ eigenvalue.
\item [${[\ubar{\textbf{I}}]_{N\times k}}$] $=[I_1,I_2,..,I_k]$: Set of first $k$ eigenvectors covers an eigen-subspace or hyperplane span$\{I_1,I_2,..,I_k\}$ of characteristic mode equation.
\item [${[\ubar{\bm{\lambda}}]_{k\times k}}$]  $=diag(\lambda_1,\lambda_2,..,\lambda_k)$: First $k$ eigenvalues of characteristic mode equation.
\item [${[\ubar{\textbf{P}}_Z]_{k\times k}}$] $=[\ubar{\textbf{I}}]^T[\textbf{Z}][\ubar{\textbf{I}}]$: Complex modal power matrix of first $k$ modes. 
\item [${[\ubar{\textbf{P}}_R]_{k\times k}}$] $= diag(P_{R1},P_{R2},...,P_{Rk})$:  Modal radiated power matrix of first $k$ modes with $P_{Rk}=[I_k]^T[\textbf{R}][I_k]$. 
\item [${[\ubar{\textbf{P}}_X]_{k\times k}}$] $= diag(P_{X1},P_{X2},...,P_{Xk})$: Modal reactive power matrix of first $k$ modes with $P_{Xk}=[I_k]^T[\textbf{X}][I_k]$. 
\item [${[\ubar{\text{E}}]}$]  $=[e_1,e_2,..e_k]^T$: Modal excitation of first $k$ modes with $e_k=[I_k]^T[\text{V}].$
\item [${[\alpha]}$] $=[\ubar{\textbf{P}}_Z]^{-1}[\ubar{\text{E}}]$: Modal contribution of first $k$ modes in total surface current $[\text{I}]\approx [\ubar{\textbf{I}}][\alpha]$.
\item [${[\textbf{Z}_{AA}]}$ and ${[\textbf{Z}_{BB}]}$]  MoM self impedance matrix of A and B.
\item [${[\textbf{Z}_{AB}]}$]  $=[\textbf{R}_{AB}]+j[\textbf{X}_{AB}]$: MoM transfer (coupling) impedance matrix between A and B.
\item [${[\ubar{\textbf{P}}_{Z,AA}]}$ and ${[\ubar{\textbf{P}}_{Z,BB}]}$] The complex (radiated and reactive) modal power matrix (diagonal) first $k$ modes of A and $l$ modes of B.
\item [${[\ubar{\textbf{P}}_{Z,AB}]}$] $=[\ubar{\textbf{P}}_{R,AB}]+j[\ubar{\textbf{P}}_{X,AB}]$: The complex (radiated and reactive) modal transfer (coupling) power matrix between first $k$ modes of A and $l$ modes of B.

\item  [${[\textbf{Z}^{uc}]}$] $=[\textbf{Z}_{AA},\textbf{0};\textbf{0},\textbf{Z}_{BB}]$: MoM impedance matrix of the uncoupled system. 
\item [${[\ubar{\bm{\lambda}}^{uc}]}$] $=[\ubar{\bm{\lambda}}_A,\bm{0};\bm{0},\ubar{\bm{\lambda}}_B]$: The set of eigenvalues of the uncoupled system $[\textbf{Z}^{uc}]$ consist of $k$ modes of $[\textbf{Z}_{AA}]$ and $l$ modes of $[\textbf{Z}_{BB}]$.
\item [${[\ubar{\textbf{I}}^{uc}]}$] $=[\ubar{\textbf{I}}_A,\bm{0};\bm{0},\ubar{\textbf{I}}_B]$: The set of eigenvectors  of the uncoupled system $[\textbf{Z}^{uc}]$ consist of $k$ modes of $[\textbf{Z}_{AA}]$ and $l$ modes of $[\textbf{Z}_{BB}]$.
\item [${[\textbf{Z}^{c}]}$] $=[\textbf{Z}_{AA},\textbf{Z}_{AB};\textbf{Z}_{BA},\textbf{Z}_{BB}]$: MoM impedance matrix of the coupled system.
\item [${[\ubar{\bm{\lambda}}^{c}]}$ and ${[\ubar{\textbf{I}}^{c}]}$]  The sets of first $k+l$ eigenvalues and eigenvectors pairs of the coupled system $[\textbf{Z}^{c}]$ consist of linear combinations $k$ modes of $[\textbf{Z}_{AA}]$ and $l$ modes of $[\textbf{Z}_{BB}]$.
\item [${[\ubar{\bm{\lambda}}^{c}_{A(B)}]}$ and ${[\ubar{\textbf{I}}^{c}_{A(B)}]}$]  The sets of first $k$ or $l$ eigenvalues and eigenvectors pairs of the coupled system $[\textbf{Z}^{c}]$ have major contributions from $k$ modes of $[\textbf{Z}_{AA}]$ or $l$ modes of $[\textbf{Z}_{BB}]$.
\item [${[\ubar{\textbf{M}}]_{(k+l)\times(k+l)}}$]  Modal coupling matrix represents modal coupling between $k$ modes of $A$ and $l$ modes $B$ with $[\ubar{\textbf{I}}^{c}]=[\ubar{\textbf{I}}^{uc}][\ubar{\textbf{M}}]$.\\
\end{IEEEdescription}

\color{black}

% \fi
\IEEEpeerreviewmaketitle

\section{Introduction}
The electromagnetic radiation and scattering phenomena of arbitrary objects are analyzed using the equivalent current vector. With the gradual demand of tightly coupled electromagnetic components within an electrically smaller real estate, it becomes challenging to determine asymptotic or analytic solution \cite{kouyoumjian1965asymptotic}–\nocite{keller1962geometrical,harrington1961time}-\cite{felsen1994radiation} of such coupled system. This inspires to explore numerical
techniques like method of moments (MoM) \cite{harrington1993field} and \cite{rao1982electromagnetic}. Although computational advancement
makes it possible to solve complex coupled structures numerically, it only provides a set of numbers with little insight into the underlying physics. It motivates to the development of various modal approaches like \cite{garbacz1965modal} and \cite{garbacz1971generalized}. For common geometries like spheres or cylinders, there exist spherical or cylindrical wave functions to analyze electromagnetic behavior. This provides the initial intuition of characteristic wave functions for arbitrarily shaped objects in \cite{garbacz1965modal} and \cite{garbacz1971generalized}. Similar to the closed-form solution of the resonating modes of commonly shaped wave-guide like structures, a mathematical representation of the resonant
scattering phenomena was initially proposed in \cite{garbacz1965modal} and \cite{garbacz1971generalized} using the eigenvectors of the object’s far-field scattering matrix. Later, the electric field integral equation (EFIE) was followed to determine the characteristic current modes through generalized eigen decomposition (GED) of the MoM impedance matrix in \cite{harrington1971theory}--\nocite{harrington1971computation}\cite{harrington1972characteristic}. Although
the electric/magnetic current-based formulation of characteristic mode analysis (CMA) was initially introduced in the seventies, the lack of suitable computing system restricted its popularity till the nineties.
With the development of computing facilities and matrix-based platforms like MATLAB \cite{MATLAB}, resurgent interests in CMA were noted for the antenna design and EMI/EMC applications \cite{bekers2006eigencurrent}--\nocite{newman1979small,bekers2009eigencurrent,cabedo2007theory,
manteuffel2016compact,capek2012method,wu2016reduction,
rothenhausler2017characteristic,sandip_temc1}\cite{gronwaldcharacteristic}.\\
It is to be noted that the computational complexities of the matrix inversion operation of conventional MoM and generalized eigendecomposition operation are the same and each follows the cubic order of the MoM impedance matrix size \cite{sghosalthesis}, unless any iterative algorithm is used. As the number of elements increases in a coupled array, the size of the  MoM impedance matrix also increases proportionately. As a consequence, the conventional CMA technique of \cite{harrington1971theory} and \cite{harrington1971computation} urges for a high computational load. This leads to the search for various alternate techniques to determine the characteristic modes of the coupled array. Initial numerical experimentation on this topic was reported in \cite{bekers2006eigencurrent} and \cite{bekers2009eigencurrent} where the modal behavior was studied in the resonating region for the linear arrays of electric and magnetic dipoles. Later, the matrix partitioning method was used to analyze two-element and three-element arrays in \cite{raines2011systematic}. For a two-element array with excitation in only one element, a different set of eigenmodes was defined as the substructure characteristic modes (SCM) in \cite{ethier2012sub}. Similarly, a lower-order matrix operation was proposed for the finite array of identical elements in \cite{lou2018analysis}. Another formulation was reported recently for the CMA of finite periodic arrays in \cite{cheng2019novel}. A rigorous matrix order reduction method was reported in \cite{kaffash2020fast} for finite symmetric arrays which provides a Toeplitz nature in the MoM impedance matrix. \blue{The behavior of mode coupling between PEC and lossy dielectric was studied in \cite{Oijala_TCCM_2019}. However, it was not discussed that the coupling of isolated modes leads to coupled characteristic modes. Characteristic modes of dipole array were studied in \cite{Lonsky_dipole_array_2018} considering half sinusoidal current in the elements. Characteristic Mode Analysis of Antenna Mutual Coupling has been discussed in \cite{Liang_Coupling_2018}, where the port-to-port admittance matrix has been used for modal analysis. However, the method proposed in \cite{Liang_Coupling_2018} does not provide a physical inside into the coupling of fundamental modes.}       

Considering the potential applications of coupled arrays in antenna and rasorber design \cite{tzanidis2012characteristic}--\nocite{ghosal2020vertical}\cite{guo2020miniaturized}, a new theory of coupled characteristic modes (TCCM) was proposed in \cite{TCCM1}. In TCCM, the modes of the isolated elements of the array are used through a weighted linear combination to construct the coupled characteristic modes of the coupled array. The main advantage of the TCCM is that it is applicable for diverse cases like in both homogeneous and non-homogeneous arrays, both symmetric and asymmetric arrays, both periodic and non-periodic arrays. Apart from the numerical benefits, it also provides a detailed physical understanding of how the isolated modes interact with each other to generate the coupled modes. It also provides a mathematical explanation behind the generation of odd (differential) and even (common) modes in the coupled array as previously also observed in \cite{cabedo2007theory}. The mutual coupling phenomena have been extensively discussed in the contexts of TCCM in \cite{TCCM2}. As shown in \cite{TCCM2}--\nocite{ghosal2019vtc,ghosal2019characteristic} \cite{ghosal2021incap}, the TCCM can play a significant role in controlling the amount of mutual coupling between two antennas. To date, the TCCM has been mostly discussed for only two-element arrays in \cite{TCCM1} and \cite{TCCM2}. As shown in \cite{TCCM1}, it requires higher-order interaction modeling to accurately determine the coupled modes for two dissimilar elements. So, the present work aims to focus on the following major extensions of the TCCM:
\begin{itemize}
\item The concept of a modal coupling matrix will be introduced.
\item It will discuss the multi-mode coupled characteristic mode.
\item It will discuss the TCCM for the coupled array of $N$ number of elements where $N\geq2$.
\item Two algorithms will be provided to calculate multi-mode coupled characteristic modes. 
\end{itemize}

\section{Eigen subspace Characteristic Mode}
\subsection{General Definition}
The method of moments equation of an antenna or scattering system can be written as \cite{harrington1993field}
\begin{equation}
[\textbf{Z}]_{N\times N}[\text{I}]_{N\times 1}=[\text{V}]_{N\times 1} \label{eq:MoM}
\end{equation}
where $\textbf{Z}$ is the discretized operator matrix of the electric field integral equation, $[\text{I}]$ is the surface current coefficient vector of $N$ discrete basis function and $[\text{V}]$ is excitation vector on $N$ discrete basis function. If the Galerkin method (i.e., both basis and testing function are identical) is applied then $[\textbf{Z}]$ is symmetric (i.e., $[\textbf{Z}]^T=[\textbf{Z}]$). The $[\textbf{Z}]$ matrix consist of two-part, written as
\begin{equation}
[\textbf{Z}]=[\textbf{R}]+j[\textbf{X}] \label{eq:Z}
\end{equation}
where the real part $[\textbf{R}]$ is responsible for the radiation and the imaginary part $[\textbf{X}]$ is responsible for the stored energy. The full eigenspace characteristic mode equation can be written as  \cite{harrington1971theory}
\begin{equation}
[\textbf{X}]_{N\times N}[\textbf{I}]_{N\times N}=[\textbf{R}]_{N\times N}[\textbf{I}]_{N\times N}[\bm{\lambda}]_{N\times N} \label{eq:fullCM}
\end{equation} 
\blue{here $[\textbf{I}]=[I_1,I_2,...,I_N]$ is the full eigen-matrix consist of $N$ eigenvectors or basis vectors $[I_n], n\in (1,2,...,N)$. The $\text{span}(\textbf{I})$ forms the complete eigenspace.} The $n^{th}$ eigenvector can be written as $[I_n]=[i_{1n},i_{2n},...i_{Nn}]^T$. The eigenvalue matrix $[\bm{\lambda}]=
  \begin{bmatrix}
    \lambda_1 & & \\
    & \ddots & \\
    & & \lambda_{N}
  \end{bmatrix}$ is a diagonal matrix consist of $N$ eigenvalue $\lambda_n, n\in (1,2,...,N)$. As $[\textbf{R}]$ and $[\textbf{X}]$ is symmetric and $[\textbf{R}]$ is positive semidefinite, the eigenvectors $[I_n], n\in (1,2,...,N)$ and eigenvalues  $\lambda_n, n\in (1,2,...,N)$ are real. 

It has been observed that few dominant (i.e., smaller magnitude of eigenvalue) characteristic modes are enough to analyze the antenna or scattering system. Therefore it is convenient to express the characteristic mode equation in a smaller set of eigenvector-based characteristic equations consisting of $k$ ($k<N$) dominant eigenvector. The subspace (span of $k$ eigenvectors) characteristic equation can be written as 
  \begin{equation}
[\textbf{X}]_{N\times N}[\ubar{\textbf{I}}]_{N\times k}=[\textbf{R}]_{N\times N}[\ubar{\textbf{I}}]_{N\times k}[\ubar{\bm{\lambda}}]_{k\times k}\label{eq:subCM}
\end{equation}
\blue{here $[\ubar{\textbf{I}}]=[I_1,I_2,...,I_k]$ is the  eigen-sub-matrix consist of $k$ eigenvector $[I_n], n\in (1,2,...,k)$.  The $\text{span}(\ubar{\textbf{I}})$ form an eigen sub-space.} The $n^{th}$ eigenvector can be written as $[I_n]=[i_{1n},i_{2n},...i_{Nn}]^T$. The eigenvalue matrix $[\ubar{\bm{\lambda}}]=
  \begin{bmatrix}
    \lambda_1 & & \\
    & \ddots & \\
    & & \lambda_{k}
  \end{bmatrix}$ is a diagonal matrix consist of $k$ eigenvalue $\lambda_n, n\in (1,2,...,k)$. 

The eigenvectors $[\ubar{\textbf{I}}]$ satisfies $[\textbf{R}]$ and $[\textbf{X}]$ orthogonality property. Properties of the generalized eigenvalue problem are given in Appendix A.  The orthogonality property of $[\ubar{\textbf{I}}]$ can be written mathematically as 
\begin{subequations}
\begin{equation}
[\ubar{\textbf{I}}]^T_{k\times N}[\textbf{R}]_{N\times N}[\ubar{\textbf{I}}]_{N\times k}={[\ubar{\textbf{P}}_R]}_{k\times k} \label{eq:PR}
\end{equation}
\begin{equation}
[\ubar{\textbf{I}}]^{T}_{k\times N}[\textbf{X}]_{N\times N}[\ubar{\textbf{I}}]_{N\times k}={[\ubar{\textbf{P}}_X]}_{k\times k}={[\ubar{\textbf{P}}_R]}_{k\times k}[\ubar{\bm{\lambda}}]_{k\times k}\label{eq:PX}
\end{equation}
\begin{equation}
[\ubar{\textbf{I}}]^{T}_{k\times N}[\textbf{Z}]_{N\times N}[\ubar{\textbf{I}}]_{N\times k}={[\ubar{\textbf{P}}_Z]}_{k\times k}={[\ubar{\textbf{P}}_R]}+j{[\ubar{\textbf{P}}_X]}\label{eq:PZ}
\end{equation}
\end{subequations}
where ${[\ubar{\textbf{P}}_R]}_{k\times k}=\begin{bmatrix}
    P_{R1} & & \\
    & \ddots & \\
    & & P_{Rk}
  \end{bmatrix}$ and ${[\ubar{\textbf{P}}_X]}_{k\times k}=\begin{bmatrix}
    P_{X1} & & \\
    & \ddots & \\
    & & P_{Xk}
  \end{bmatrix}$ are diagonal matrix with diagonal element $P_{Rn}=[I_n]^T[\textbf{R}][I_n]$ and $P_{Xn}=[I_n]^T[\textbf{X}][I_n]$. $P_{Rn}$ and $P_{Xn}$ are related modal radiation power and reactive power of the $n^{th}$ mode. The eigenvalues $\lambda_n$ are the ratio of reactive power and radiation power (i.e., $\lambda_n=P_{Xn}/P_{Rn}$). In the other words, the reduced eigenvalue matrix $[\ubar{\bm{\lambda}}]$ can be written as 
  \begin{equation}
  [\ubar{\bm{\lambda}}]_{k\times k}={[\ubar{\textbf{P}}_R]}^{-1}_{k\times k}{[\ubar{\textbf{P}}_X]}_{k\times k}=\begin{bmatrix}
    \frac{P_{X1}}{P_{R1}} & & \\
    & \ddots & \\
    & & \frac{P_{Xk}}{P_{Rk}}
  \end{bmatrix} \label{eq:lam}
  \end{equation}
The solution of $[\text{I}]$ in \eqref{eq:MoM} can be approximated as weighted sum of first $k$ dominant eigenvectors $[\ubar{\textbf{I}}]_{N\times k}$ as
\begin{equation}
[\text{I}]\approx [\ubar{\textbf{I}}]_{N\times k}[\alpha]_{k\times 1}\label{eq:Iapprox}
\end{equation}
where $[\alpha]=[\alpha_1,\alpha_2,...,\alpha_k]^T$ is weighting coefficient vector with $\alpha_n$ is the weighting coefficient of the $n^{th}$ mode. One can rewrite \eqref{eq:MoM} by replacing $[\text{I}]$ as in \eqref{eq:Iapprox} as
\begin{equation}
[\textbf{Z}]_{N\times N}[\ubar{\textbf{I}}]_{N\times k}[\alpha]_{k\times 1}=[\text{V}]_{N\times 1}.\label{eq:MoMapprox}
\end{equation}
Taking inner product with  $[\ubar{\textbf{I}}]$ to the both side of \eqref{eq:MoMapprox} we have
\begin{equation}
[\ubar{\textbf{I}}]^T_{k\times N}[\textbf{Z}]_{N\times N}[\ubar{\textbf{I}}]_{N\times k}[\alpha]_{k\times 1}=[\ubar{\textbf{I}}]^T_{k\times N}[\text{V}]_{N\times 1}=[\ubar{\text{E}}]_{k\times1}.\label{eq:MoMapprox1}
\end{equation}
where $[\ubar{\text{E}}]=[e_1,e_2,...e_k]^T$ is the excitation dependent coefficient with $e_n=[I_n]^T[\text{V}], n \in (1,2,...k)$. The equation \eqref{eq:MoMapprox1} can be solved for $[\alpha]$ as 
\begin{equation}
[\alpha]_{k\times 1}={[\ubar{\textbf{P}}_Z]}^{-1}_{k\times k}[E]_{k\times1}\label{eq:alpha}
\end{equation} 
where $n^{th}$ weighting coefficient can be written as
\begin{equation}
\alpha_n=\frac{e_n}{P_{Zn}}=\frac{[I_n]^T[V]}{P_{Rn}+jP_{Xn}}=\frac{[I_n]^T[V]}{P_{Rn}(1+j\lambda_n)}\label{eq:alphan}
\end{equation}
This completes our understanding of eigen-subspace of characteristic mode. Next we will understand coupled mode system.
\subsection{Coupled System mode}
Let's consider a two-antenna (or scatterer) coupling system consisting of antenna $A$ and antenna $B$. The EFIE operator matrix of antenna $A$ discretized into $M$ dimension and antenna $B$ discretized into $N$ dimension. The antenna $A$ has $k$ dominant characteristic mode and antenna $B$ has $l$ dominant characteristic mode. The subspace characteristic mode equation of antenna $A$ and antenna $B$ can be written as
\begin{subequations}
\begin{equation}
[\textbf{X}_{AA}]_{M\times M}[\ubar{\textbf{I}}_A]_{M\times k}=[\textbf{R}_{AA}]_{M\times M}[\ubar{\textbf{I}}_A]_{M\times k}[\ubar{\bm{\lambda}}_A]_{k\times k}\label{eq:CM_a}
\end{equation}
\begin{equation}
[\textbf{X}_{BB}]_{N\times N}[\ubar{\textbf{I}}_B]_{N\times l}=[\textbf{R}_{BB}]_{N\times N}[\ubar{\textbf{I}}_B]_{N\times l}[\ubar{\bm{\lambda}}_B]_{l\times l}\label{eq:CM_b}
\end{equation}
\end{subequations}
The first $k+l$ dominant character mode equations of the uncoupled antenna system can be written using \eqref{eq:CM_a} and \eqref{eq:CM_b} as
\begin{align}
&
\begin{bmatrix}
[\textbf{X}_{AA}] & [\bm{0}]_{M\times N}\\
[\bm{0}]_{N\times M} & [\textbf{X}_{BB}]
\end{bmatrix}
\begin{bmatrix}
[\ubar{\textbf{I}}_A] &[\bm{0}]_{M\times l}\\
[\bm{0}]_{N\times k} & [\ubar{\textbf{I}}_B]
\end{bmatrix}=\nonumber\\
&\begin{bmatrix}
[\textbf{R}_{AA}] & [\bm{0}]_{M\times N}\\
[\bm{0}]_{N\times M} & [\textbf{R}_{BB}]
\end{bmatrix}
\begin{bmatrix}
[\ubar{\textbf{I}}_A] &[\bm{0}]_{M\times l}\\
[\bm{0}]_{N\times k} & [\ubar{\textbf{I}}_B]
\end{bmatrix}
\begin{bmatrix}
[\ubar{\bm{\lambda}}_A] &[\bm{0}]_{k\times l}\\
[\bm{0}]_{l\times k} & [\ubar{\bm{\lambda}}_B]
\end{bmatrix}\label{eq:CM_uc}
\end{align} 
The first $k+l$ dominant characteristic mode equations of the coupled antenna system can be written as 
 \begin{align}
&
\begin{bmatrix}
[\textbf{X}_{AA}] & [\textbf{X}_{AB}]\\
[\textbf{X}_{BA}] & [\textbf{X}_{BB}]
\end{bmatrix}
\begin{bmatrix}
[\ubar{\textbf{I}}^{cA}_A] &[\ubar{\textbf{I}}^{cB}_A]\\
[\ubar{\textbf{I}}^{cA}_B] & [\ubar{\textbf{I}}^{cB}_B]
\end{bmatrix}=\nonumber\\
&\begin{bmatrix}
[\textbf{R}_{AA}] & [\textbf{R}_{AB}]\\
[\textbf{R}_{BA}] & [\textbf{R}_{BB}]
\end{bmatrix}
\begin{bmatrix}
[\ubar{\textbf{I}}^{cA}_A] &[\ubar{\textbf{I}}^{cB}_A]\\
[\ubar{\textbf{I}}^{cA}_B] & [\ubar{\textbf{I}}^{cB}_B]
\end{bmatrix}
\begin{bmatrix}
[\ubar{\bm{\lambda}}^{cA}] &[\bm{0}]_{k\times l}\\
[\bm{0}]_{l\times k} & [\ubar{\bm{\lambda}}^{cB}]
\end{bmatrix}.\label{eq:CM_c}
\end{align} 
Here $[\textbf{Z}_{AB}]_{M\times N}=[\textbf{R}_{AB}]+j[\textbf{X}_{AB}]$ is the mutual impedance matrix, and $[\textbf{Z}_{BA}]_{N\times M}=[\textbf{Z}_{AB}]^T=[\textbf{R}_{AB}]^T+j[\textbf{X}_{AB}]^T$. The coupled mode space $[\ubar{\textbf{I}}^{cA}]_{M+N\times k}=[[\ubar{\textbf{I}}^{cA}_A]^T,[\ubar{\textbf{I}}^{cA}_B]^T]^T$ 
is the $A$ dominant (i.e., $max(|[I^{cA}_{An}]|)>max(|[I^{cA}_{Bn}]|)$) 
eigen subspace and $[\ubar{\bm{\lambda}}^{cA}]$ is the $A$ dominant sub eigen value matrix of the coupled system. Similarly $[\ubar{\textbf{I}}^{cB}]_{M+N\times l}=[[\ubar{\textbf{I}}^{cB}_A]^T,[\ubar{\textbf{I}}^{cB}_B]^T]^T$ is the $B$ dominant eigen subspace and $[\ubar{\bm{\lambda}}^{cB}]$ is the $B$ dominant sub eigen value matrix of the coupled system.

\section{Theory of Coupled Characteristic Mode}

The coupled characteristic modes are linearly related to the isolated characteristic mode \cite{TCCM1}. In the special theory of coupled characteristic mode \cite{TCCM1}-\cite{TCCM2}, we have considered one-to-one and two-two mode interaction between isolated modes. One-to-one modal interaction approximation is accurate enough in $1D$ geometry of sub-wavelength dimension but less accurate in $2D$ geometry. Similarly, two to two modal interaction approximation is good for $2D$ geometry but may not explain the modal coupling of $3D$ structure.   In this section, we will consider many to many modes of interaction between isolated modes. The general theory of coupled characteristic mode (multi-mode interaction) can 
be summarized as the following Theorem given below.
 \begin{theorem}\label{th1}
 : Let $[\ubar{\textbf{I}}_A]_{M\times k}$ are the first $k$ dominant characteristic modes of antenna A and $[\ubar{\textbf{I}}_B]_{N\times l}$ are the first $l$ dominant characteristic modes of antenna B. Then there exist $k+l$ coupled characteristic modes $[\ubar{\textbf{I}}^{c}]_{M+N\times k+l}$ and eigenvalue $[\ubar{\bm{\lambda}}^{c}]$, which are linearly related to the $[\ubar{\textbf{I}}_A]_{M\times k}$ and $[\ubar{\textbf{I}}_B]_{N\times l}$ using mode coupling matrix $[\ubar{\textbf{M}}]_{(k+l)\times (k+l)}$. Where
 \begin{subequations}
 \begin{equation}
 [\ubar{\textbf{M}}]=
 \begin{bmatrix}
 [\ubar{\textbf{M}}^{A}_{A}]_{k\times k} & [\ubar{\textbf{M}}^{B}_{A}]_{k\times l}\\
 [\ubar{\textbf{M}}^{A}_{B}]_{l\times k} & [\ubar{\textbf{M}}^{B}_{B}]_{l\times l}
 \end{bmatrix}\label{eq:M}
 \end{equation}
 \begin{equation}
 [\ubar{\textbf{I}}^{c}]=[\ubar{\textbf{I}}^{uc}][\ubar{\textbf{M}}]=
 \begin{bmatrix}
 [\ubar{\textbf{I}}_A][\ubar{\textbf{M}}^{A}_{A}] & [\ubar{\textbf{I}}_A][\ubar{\textbf{M}}^{B}_{A}]\\
 [\ubar{\textbf{I}}_B][\ubar{\textbf{M}}^{A}_{B}] & [\ubar{\textbf{I}}_B][\ubar{\textbf{M}}^{B}_{B}]
 \end{bmatrix}.\label{eq:Ic}
 \end{equation}
 \end{subequations}
 The mode coupling matrix $[{\textbf{M}}]$ and $[\ubar{\bm{\lambda}}^{c}]$ are solution of the following generalized eigen equation of dimension $(k+l)\times (k+l)$
 
 \begin{align}
&
\begin{bmatrix}
[\ubar{\textbf{P}}_{X,AA}] & [\ubar{\textbf{P}}_{X,AB}]\\
[\ubar{\textbf{P}}_{X,BA}] & [\ubar{\textbf{P}}_{X,BB}]
\end{bmatrix}
\begin{bmatrix}
 [\ubar{\textbf{M}}^{A}_{A}] & [\ubar{\textbf{M}}^{B}_{A}]\\
 [\ubar{\textbf{M}}^{A}_{B}] & [\ubar{\textbf{M}}^{B}]
 \end{bmatrix}=\nonumber\\
&\begin{bmatrix}
[\ubar{\textbf{P}}_{R,AA}] & [\ubar{\textbf{P}}_{R,AB}]\\
[\ubar{\textbf{P}}_{R,BA}] & [\ubar{\textbf{P}}_{R,BB}]
\end{bmatrix}
\begin{bmatrix}
 [\ubar{\textbf{M}}^{A}_{A}] & [\ubar{\textbf{M}}^{B}_{A}]\\
 [\ubar{\textbf{M}}^{A}_{B}] & [\ubar{\textbf{M}}^{B}]
 \end{bmatrix}
\begin{bmatrix}
[\ubar{\bm{\lambda}}^{cA}] &[\bm{0}]\\
[\bm{0}] & [\ubar{\bm{\lambda}}^{cB}]
\end{bmatrix}.\label{eq:CCM}
\end{align} 
Here 
\begin{subequations}
\allowdisplaybreaks
    \begin{align}
      [\ubar{\textbf{P}}_{R,AA}]_{k\times k}=[\ubar{\textbf{I}}_A]^T[\textbf{R}_{AA}][\ubar{\textbf{I}}_A]\\
[\ubar{\textbf{P}}_{R,BB}]_{l\times l}=[\ubar{\textbf{I}}_B]^T[\textbf{R}_{BB}][\ubar{\textbf{I}}_B]\\ 
[\ubar{\textbf{P}}_{R,AB}]_{k\times l}=[\ubar{\textbf{I}}_A]^T[\textbf{R}_{AB}][\ubar{\textbf{I}}_B]\\ 
[\ubar{\textbf{P}}_{R,BA}]_{l\times k}=[\ubar{\textbf{P}}_{R,AB}]^T\\ 
[\ubar{\textbf{P}}_{X,AA}]=[\ubar{\textbf{P}}_{R,AA}][\ubar{\bm{\lambda}}_{A}]\\
[\ubar{\textbf{P}}_{X,BB}]=[\ubar{\textbf{P}}_{R,BB}][\ubar{\bm{\lambda}}_{B}]\\
[\ubar{\textbf{P}}_{X,AB}]_{k\times l}=[\ubar{\textbf{I}}_A]^T[\textbf{X}_{AB}][\ubar{\textbf{I}}_B]\\ 
[\ubar{\textbf{P}}_{X,BA}]_{l\times k}=[\ubar{\textbf{P}}_{X,AB}]^T.  
    \end{align}
\end{subequations}

 \end{theorem}
 \begin{proof}
 Using \eqref{eq:CM_uc} the first $k+l$ dominant characteristic mode of uncoupled antenna array A and B can be expressed as
 \begin{subequations}
 \begin{equation}
[\ubar{\textbf{I}}^{uc}]_{(M+N)\times(k+l)}= \begin{bmatrix}
[\ubar{\textbf{I}}_A] &[\bm{0}]\\
[\bm{0}] & [\ubar{\textbf{I}}_B]
\end{bmatrix}
 \end{equation}
 \begin{equation}
 [\ubar{\bm{\lambda}}^{uc}]_{(k+l)\times(k+l)}=\begin{bmatrix}
[\ubar{\bm{\lambda}}_A] &[\bm{0}]\\
[\bm{0}] & [\ubar{\bm{\lambda}}_B]
\end{bmatrix}
 \end{equation}
 \end{subequations}
 Following the idea that the fundamental modes of A are coupled to the fundamental modes of B, and there is no (or negligible) coupling between the dominant modes and higher order modes, one can express the  $(k+l)$ dominant coupled characteristic mode $[\ubar{\textbf{I}}^{c}]$ as linear combination of first $k+l$ dominant uncoupled characteristic mode  $[\ubar{\textbf{I}}^{uc}]$ using coupling matrix $[\ubar{\textbf{M}}]$. So we can write
 \begin{equation}
 [\ubar{\textbf{I}}^{c}]=[\ubar{\textbf{I}}^{uc}][\ubar{\textbf{M}}].
 \end{equation}
 The characteristic mode equation \eqref{eq:CM_c} of the coupled system can be expressed as
 \begin{equation}
 [\textbf{X}^c][\ubar{\textbf{I}}^{uc}][\ubar{\textbf{M}}]= [\textbf{R}^c][\ubar{\textbf{I}}^{uc}][\ubar{\textbf{M}}][\ubar{\bm{\lambda}}^{c}].\label{eq:CM_c1}
 \end{equation}
 Taking the inner product on both side of \eqref{eq:CM_c1}  with $[\ubar{\textbf{I}}^{uc}]$, we have
 \begin{equation}
 [\ubar{\textbf{I}}^{uc}]^T[\textbf{X}^c][\ubar{\textbf{I}}^{uc}][\ubar{\textbf{M}}]= [\ubar{\textbf{I}}^{uc}]^T[\textbf{R}^c][\ubar{\textbf{I}}^{uc}][\ubar{\textbf{M}}][\ubar{\bm{\lambda}}^{c}].\label{eq:CM_c2}
 \end{equation}
 The equation \eqref{eq:CM_c2} can be expressed as
 \begin{equation}
 [\ubar{\textbf{P}}_{Xc}][\ubar{\textbf{M}}]= [\ubar{\textbf{P}}_{Rc}][\ubar{\textbf{M}}][\ubar{\bm{\lambda}}^{c}]. \label{eq:CM_c3}
 \end{equation}
 where
 \begin{subequations}
 \begin{align}
 [\ubar{\textbf{P}}_{Xc}]=&
 \begin{bmatrix}
[\ubar{\textbf{I}}_A]^T &[\bm{0}]\\
[\bm{0}] & [\ubar{\textbf{I}}_B]^T
\end{bmatrix}
\begin{bmatrix}
[\textbf{X}_{AA}] & [\textbf{X}_{AB}]\\
[\textbf{X}_{BA}] & [\textbf{X}_{BB}]
\end{bmatrix}
\begin{bmatrix}
[\ubar{\textbf{I}}_A] &[\bm{0}]\\
[\bm{0}] & [\ubar{\textbf{I}}_B]
\end{bmatrix}\nonumber\\
=&
\begin{bmatrix}
[\ubar{\textbf{I}}_A]^T[\textbf{X}_{AA}][\ubar{\textbf{I}}_A] & [\ubar{\textbf{I}}_A]^T[\textbf{X}_{AB}][\ubar{\textbf{I}}_B]\\
[\ubar{\textbf{I}}_B]^T[\textbf{X}_{BA}][\ubar{\textbf{I}}_A] & [\ubar{\textbf{I}}_B]^T[\textbf{X}_{BB}][\ubar{\textbf{I}}_B]
\end{bmatrix}\label{eq:Px}
 \end{align}
 \begin{align}
 [\ubar{\textbf{P}}_{Rc}]=&
 \begin{bmatrix}
[\ubar{\textbf{I}}_A]^T &[\bm{0}]\\
[\bm{0}] & [\ubar{\textbf{I}}_B]^T
\end{bmatrix}
\begin{bmatrix}
[\textbf{R}_{AA}] & [\textbf{R}_{AB}]\\
[\textbf{R}_{BA}] & [\textbf{R}_{BB}]
\end{bmatrix}
\begin{bmatrix}
[\ubar{\textbf{I}}_A] &[\bm{0}]\\
[\bm{0}] & [\ubar{\textbf{I}}_B]
\end{bmatrix}\nonumber\\
=&
\begin{bmatrix}
[\ubar{\textbf{I}}_A]^T[\textbf{R}_{AA}][\ubar{\textbf{I}}_A] & [\ubar{\textbf{I}}_A]^T[\textbf{R}_{AB}][\ubar{\textbf{I}}_B]\\
[\ubar{\textbf{I}}_B]^T[\textbf{R}_{BA}][\ubar{\textbf{I}}_A] & [\ubar{\textbf{I}}_B]^T[\textbf{R}_{BB}][\ubar{\textbf{I}}_B]
\end{bmatrix}\label{eq:Pr}
 \end{align}
 \end{subequations}
Equation \eqref{eq:CCM} is identical with \eqref{eq:CM_c3}, \eqref{eq:Pr} and \eqref{eq:Px}. Hence the theorem is proved.
 \end{proof}
 To elaborate the Theorem consider an example of two-to-two mode coupling between antenna A and antenna B. The fundamental isolated mode pairs are selected based on the following criterion
 \begin{subequations}
 \begin{equation}
 \ubar{\textbf{P}}_{R/XAB}(i,j)\neq 0
 \end{equation}
 where
 \begin{equation}
 [\ubar{\textbf{P}}_{R/XAB}]=max([|\ubar{\textbf{P}}_{RAB}|],[|\ubar{\textbf{P}}_{XAB}|]).
 \end{equation}
 \end{subequations}
 Let mode (1, 3) of antenna A be coupled with mode (2, 4) of antenna B. Then 
 \begin{subequations}
 \begin{equation}
 [\ubar{\textbf{P}}_{Rc}]=
 \begin{bmatrix}
 P_{R1}^{A}&0&P_{R12}^{AB}&P_{R14}^{AB}\\
 0&P_{R3}^{A}&P_{R32}^{AB}&P_{R34}^{AB}\\
 P_{R12}^{AB}& P_{R32}^{AB}&P_{R2}^{B}&0\\
 P_{R14}^{AB}& P_{R34}^{AB}&0&P_{R4}^{B}
 \end{bmatrix}
 \end{equation}
 \begin{equation}
 [\ubar{\textbf{P}}_{Xc}]=
 \begin{bmatrix}
  P_{X1}^{A}&0&P_{X12}^{AB}&P_{X14}^{AB}\\
 0&P_{X3}^{A}&P_{X32}^{AB}&P_{X34}^{AB}\\
 P_{X12}^{AB}& P_{X32}^{AB}&P_{X2}^{B}&0\\
 P_{X14}^{AB}& P_{X34}^{AB}&0&P_{X4}^{B}
 \end{bmatrix}
 \end{equation}
 here
 \begin{equation}
 P_{Ri}^{A}=[I_{iA}]^T[\textbf{R}_{AA}][I_{iA}];\; P_{Rij}^{AB}=[I_{iA}]^T[\textbf{R}_{AB}][I_{jB}]
 \end{equation}
  \begin{equation}
 P_{Xi}^{A}=[I_{iA}]^T[\textbf{X}_{AA}][I_{iA}];\; P_{Xij}^{AB}=[I_{iA}]^T[\textbf{X}_{AB}][I_{jB}]
 \end{equation}
 and
 \begin{equation}
 max(| P_{Rij}^{AB}|,| P_{Xij}^{AB}|)\neq 0.
 \end{equation}
 \end{subequations}
 The coupling matrix between modes (1, 3) of A and (2, 4) of B can be defined as
 \begin{equation}
 [\textbf{M}_{(A13)(B24)}]=
 \begin{bmatrix}
 1&m^{A3}_{A1}&m^{B2}_{A1}&m^{B4}_{A1}\\
 m^{A1}_{A3}&1&m^{B2}_{A3}&m^{B4}_{A3}\\
 m^{A1}_{B2}&m^{A3}_{B2}&1&m^{B4}_{A2}\\
 m^{A1}_{B4}&m^{A3}_{B4}&m^{B2}_{B4}&1
 \end{bmatrix}
 \end{equation}
 Here we have considered that 1 is the maximum value in the coupling matrix. The first column represents the coupled mode produced due to the maximum contribution from mode-1 of A. Similarly, the third column represents the coupled mode produced due to the maximum contribution from mode-2 of B.
  
 We have discussed the theory of multi-mode coupled characteristic mode of two-element arrays. For the practical computation of multi-mode coupled characteristic mode, we need to develop a computer algorithm. Based on the proposed theory, multi-mode coupled characteristic mode Algorithm-\ref{alg1} is developed.  Algorithm-\ref{alg1} can easily be incorporated into existing characteristic mode solvers.  
 \begin{algorithm}
\caption{Multi-mode Coupling of Two Element Array}
\begin{algorithmic} 
\Require $[\textbf{Z}_{AA}]$, $[\textbf{Z}_{BB}]$, $[\textbf{Z}_{AB}]$, $k$, $l$
\Ensure $[\ubar{\bm{\lambda}}^{c}]$, $[\ubar{\textbf{M}}]$, $[\ubar{\textbf{I}}^{uc}]$ and $[\ubar{\textbf{I}}^{c}]$. 
\State $M=\text{len}(\textbf{Z}_{AA})$; $N=\text{len}(\textbf{Z}_{BB})$
\State $\textbf{R}_{AA}=\text{real}(\textbf{Z}_{AA})$; $\textbf{X}_{AA}=\text{imag}(\textbf{Z}_{AA})$
\State $\textbf{R}_{BB}=\text{real}(\textbf{Z}_{BB})$; $\textbf{X}_{BB}=\text{imag}(\textbf{Z}_{BB})$
\State $\textbf{Z}_{BA}=\textbf{Z}_{AB}^T$
\State $[{\textbf{I}}_{A},{\bm{\lambda}}_A]=\text{eig}(\textbf{X}_{AA},\textbf{R}_{AA})$
\State $[{\textbf{I}}_{B},{\bm{\lambda}}_B]=\text{eig}(\textbf{X}_{BB},\textbf{R}_{BB})$
\State $[\ubar{\textbf{I}}_{A}]={\textbf{I}}_{A}(:,1:k)$
\State $[\ubar{\textbf{I}}_{B}]={\textbf{I}}_{A}(:,1:l)$
\State $\textbf{Z}^c=[\textbf{Z}_{AA},\textbf{Z}_{AB};\textbf{Z}_{BA},\textbf{Z}_{BB}]$
\State $\ubar{\textbf{I}}^{uc}=[\ubar{\textbf{I}}_{A},\textbf{0};\textbf{0},\ubar{\textbf{I}}_{B}]$
\State $\ubar{\textbf{P}}_{Zc}=[\ubar{\textbf{I}}^{uc}]^T[\textbf{Z}^c][\ubar{\textbf{I}}^{uc}]$ 
\State $\ubar{\textbf{P}}_{Rc}=\text{real}(\ubar{\textbf{P}}_{Zc})$; $\ubar{\textbf{P}}_{Xc}=\text{imag}(\ubar{\textbf{P}}_{Zc})$
\State $[\ubar{\textbf{M}},\ubar{\bm{\lambda}}^{c}]=\text{eig}(\ubar{\textbf{P}}_{Xc},\ubar{\textbf{P}}_{Rc})$
\State $\ubar{\textbf{I}}^{c}=[\ubar{\textbf{I}}^{uc}][\ubar{\textbf{M}}]$
\end{algorithmic}
\label{alg1}
\end{algorithm}

 The above theorem describe multi-mode coupling between two antenna or scatterer arrays. We can further generalize the theorem for the $N$ element array. For the $N$ element array, the coupled characteristic mode can be predicted from individual antenna modes and coupling matrix using the following theorem 
 \begin{theorem}\label{th2}
 In an $N$ element array each antenna has $K_n$ characteristic mode (i.e., $[{\textbf{I}_n}]_{K_n\times K_n}$) and $k_n$ ($n\in (1,2...N)$)  dominant modes (i.e., $[\ubar{\textbf{I}}_n]_{K_n\times k_n}$). Then $k=\sum_{n=1}^{N} k_n$ dominant modes of coupled antenna (i.e., $[\ubar{\textbf{I}}^c]_{K\times k}$, $K=\sum_{n=1}^{N} K_n$) can be predicted as linear combination of individual mode $[\ubar{\textbf{I}}_n]_{K_n\times k_n}$ and coupling matrix $[\ubar{\textbf{M}}]$. Where 
 \begin{subequations}
 \begin{equation}
 [\ubar{\textbf{M}}]=
 \begin{bmatrix}
 [\ubar{\textbf{M}}^{1}_{1}]& \hdots &[\ubar{\textbf{M}}^{N}_{1}]\\
 \vdots & \ddots & \vdots\\
 [\ubar{\textbf{M}}^{1}_{N}]& \hdots&[\ubar{\textbf{M}}^{N}_{N}]
 \end{bmatrix}
 \end{equation}
 \begin{equation}
 [\ubar{\textbf{I}}^c]=
 \begin{bmatrix}
 [\ubar{\textbf{I}}_1][\ubar{\textbf{M}}^{1}_{1}]& \hdots &[\ubar{\textbf{I}}_1][\ubar{\textbf{M}}^{N}_{1}]\\
 \vdots & \ddots & \vdots\\
 [\ubar{\textbf{I}}_N][\ubar{\textbf{M}}^{1}_{N}]& \hdots&[\ubar{\textbf{I}}_N][\ubar{\textbf{M}}^{N}_{N}]
 \end{bmatrix}
 \end{equation}
 \end{subequations}
 The mode coupling matrix $[\ubar{\textbf{M}}]$ and and eigenvalues $[\ubar{\bm{\lambda}}^{c}]$ are solution of the following generalized eigen equation of dimension $k\times k$
 \begin{equation}
 [\ubar{\textbf{P}}_{Xc}][\ubar{\textbf{M}}]= [\ubar{\textbf{P}}_{Rc}][\ubar{\textbf{M}}][\ubar{\bm{\lambda}}^{c}].
 \end{equation}
 where $[\ubar{\textbf{P}}_{Xc}]$ and $[\ubar{\textbf{P}}_{Rc}]$ is consist of $N\times N$ sub block matrix. The sub-block matrices are given as
 \begin{subequations}
 \begin{equation}
 [\ubar{\textbf{P}}_{R,ij}]=[\ubar{\textbf{I}}_i]^T[\textbf{R}_{ij}][\ubar{\textbf{I}}_j]
\end{equation} 
\begin{equation}
 [\ubar{\textbf{P}}_{X,ij}]=[\ubar{\textbf{I}}_i]^T[\textbf{X}_{ij}][\ubar{\textbf{I}}_j]
\end{equation} 
 \end{subequations}
 with $i\in(1,2..N)$ and $j\in (1,2...N)$
\end{theorem} 

\begin{proof}
Using uncoupled mode theory one can write first $k=k_1+k_2+...+k_N$ dominant modes of  $N$ isolated antenna system having $[\ubar{\textbf{I}}_1]$, $[\ubar{\textbf{I}}_2]$,..., $[\ubar{\textbf{I}}_N]$ dominant modes of each antenna with dimension $K_n\times k_n$, as

\begin{align}
[\ubar{\textbf{I}}^{uc}]_{K\times k}=\begin{bmatrix}
[\ubar{\textbf{I}}_1] & & &\\
& [\ubar{\textbf{I}}_2]& &\\
   & & \ddots & \\
    & & & [\ubar{\textbf{I}}_N]
\end{bmatrix}
\end{align}
\begin{align}
[\ubar{\bm{\lambda}}^{uc}]_{k\times k}=\begin{bmatrix}
[\ubar{\bm{\lambda}}_1] & & &\\
& [\ubar{\bm{\lambda}}_2]& &\\
   & & \ddots & \\
    & & & [\ubar{\bm{\lambda}}_N]
\end{bmatrix}
\end{align}
As we have discussed two two-element coupled characteristic modes are linear combinations of uncoupled characteristic modes, the same is true for N elements coupled characteristic mode. Therefore, the first $k=k_1+k_2+..+k_N$ coupled characteristic subspace of N body coupled characteristic modes can be represented as 
\begin{align}
[\ubar{\textbf{I}}^{c}]_{K\times k}=[\ubar{\textbf{I}}^{uc}]_{K\times k}[\ubar{\textbf{M}}]_{k\times k}
\end{align}  
linear combination of uncoupled characteristic modes. Where $[\ubar{\textbf{M}}]$ is coupling matrix. The subspace coupled characteristic mode equation of the N-body problem can be written as
\begin{align}
[\textbf{X}^c][\ubar{\textbf{I}}^{uc}][{\textbf{M}}]= [\textbf{R}^c][\ubar{\textbf{I}}^{uc}][{\textbf{M}}][\ubar{\bm{\lambda}}^{c}],\label{eq:CCMn}
\end{align}
where
\begin{align}
[\textbf{Z}^{c}]=[\textbf{R}^{c}]+j[\textbf{X}^{c}]=\begin{bmatrix}
[\textbf{Z}_{11}] & [\textbf{Z}_{12}] & \hdots & [\textbf{Z}_{1N}]\\
[\textbf{Z}_{21}] & [\textbf{Z}_{22}] & \hdots & [\textbf{Z}_{2N}\\
\vdots & \vdots & \ddots & \vdots\\
[\textbf{Z}_{N1}] & [\textbf{Z}_{N2}] & \hdots & [\textbf{Z}_{NN}\\
\end{bmatrix}
\end{align}
is impedance matrix of the coupled N-body system with  $[\textbf{Z}_{ii}]=[\textbf{R}_{ii}]+j[\textbf{X}_{ii}]$ is self impedance matrix of $i^{th}$ body and $[\textbf{Z}_{ij}]=[\textbf{R}_{ij}]+j[\textbf{X}_{ij}]$ is mutual impedance matrix between $i^{th}$ and $j^{th}$ bodies. Please note that $[\textbf{Z}_c]$ is reciprocal (i.e., $[\textbf{Z}^c]^T=[\textbf{Z}^c]$) with $[\textbf{Z}_{ij}]=[\textbf{Z}_{ji}]^T$.

Taking inner-product with respect to $[\ubar{\textbf{I}}^{uc}]$ on both side of \eqref{eq:CCMn} leads to
\begin{align}
[\ubar{\textbf{I}}^{uc}]^T[\textbf{X}^c][\ubar{\textbf{I}}^{uc}][{\textbf{M}}]= [\ubar{\textbf{I}}^{uc}]^T[\textbf{R}^c][\ubar{\textbf{I}}^{uc}][{\textbf{M}}][\ubar{\bm{\lambda}}^{c}].\label{eq:CCMn1}
\end{align}
After matrix multiplication, the equation \eqref{eq:CCMn1} simplifies to a generalized eigenvalue problem 
\begin{align}
[\ubar{\textbf{P}}_{Xc}][\ubar{\textbf{M}}]= [\ubar{\textbf{P}}_{Rc}][\ubar{\textbf{M}}][\ubar{\bm{\lambda}}^{c}]\label{eq:CCMn2}
\end{align} 
where
\begin{align}
[\ubar{\textbf{P}}_{Zc}] =[\ubar{\textbf{P}}_{Rc}]+j[\ubar{\textbf{P}}_{Xc}]=\begin{bmatrix}
[\ubar{\textbf{P}}_{Z,11}] & [\ubar{\textbf{P}}_{Z,12}] & \hdots & [\ubar{\textbf{P}}_{Z,1N}]\\
[\ubar{\textbf{P}}_{Z,21}] & [\ubar{\textbf{P}}_{Z,22}] & \hdots & [\ubar{\textbf{P}}_{Z,2N}\\
\vdots & \vdots & \ddots & \vdots\\
[\ubar{\textbf{P}}_{Z,N1}] & [\ubar{\textbf{P}}_{Z,N2}] & \hdots & [\ubar{\textbf{P}}_{Z,NN}\\
\end{bmatrix}
\end{align}
is $k\times k$ complex power matrix. The diagonal submatrices $[\ubar{\textbf{P}}_{R,ii}]_{k_i\times k_i}=[\ubar{\textbf{I}}_i]^T[\textbf{R}_{ii}][\ubar{\textbf{I}}_i]=\begin{bmatrix}
    P_{R1} & & \\
    & \ddots & \\
    & & P_{Rk_i}
  \end{bmatrix}$ are individually diagonal square matrices. On other hands non-diagonal blocks are rectangular $[\ubar{\textbf{P}}_{R,ij}]_{k_i\times k_j}=[\ubar{\textbf{I}}_i]^T[\textbf{R}_{ij}][\ubar{\textbf{I}}_j]=\begin{bmatrix}
    P_{R11} &\hdots & P_{R1k_i}\\
    \vdots& \ddots & \vdots \\
    P_{Rk_j1}& \hdots& P_{Rk_ik_j}
  \end{bmatrix}$ are dense rectangular matrix.  
\end{proof}
The theory of multi-mode multi-element coupling of characteristic modes has been discussed in Theorem-\ref{th2}. The theorem is a little bit difficult to understand as compared to two-element multi-mode coupling. An algorithm (Algorithm-\ref{alg2}) is developed for multi-element multi-mode coupling. This algorithm returns $k=\sum_{i=1}^N k_i$ eigenvalues $[\ubar{\bm{\lambda}}^{c}]$ and eigenvector $[\ubar{\textbf{I}}^{c}]$  coupled characteristic mode. It also returns modal coupling matrix $[\ubar{\textbf{M}}]$, which is a measure of coupling between characteristic modes.   

\begin{algorithm}
\caption{Multi-mode Coupling of Multi-Element Array}
\begin{algorithmic} 
\Require $N$, [$[\textbf{Z}_{11}]$, $[\textbf{Z}_{22}]$, $\hdots$, $[\textbf{Z}_{NN}]$], \\
 $[k_1,  k_2,..., k_N]$,\\
 $[[\textbf{Z}_{12}],[\textbf{Z}_{13},\hdots,[\textbf{Z}_{1N}]]$,\\
 $[[\textbf{Z}_{23}],[\textbf{Z}_{24}],\hdots,[\textbf{Z}_{2N}]]$,\\
 $\vdots$\\
 $[[\textbf{Z}_{(N-2)(N-1)}],[\textbf{Z}_{(N-2)N}]]$,\\
 $[\textbf{Z}_{(N-1)N}]$
%[$[\textbf{Z}_{12}]$, $\hdots$, $[\textbf{Z}_{1N}]$], [$[\textbf{Z}_{23}]$, $\hdots$, $[\textbf{Z}_{2N}]$], $\hdots$, $[\textbf{Z}_{(N-2)(N-1)}]$, $[\textbf{Z}_{(N-2)N}]$, $[\textbf{Z}_{(N-1)N}]$.
\Ensure $[\ubar{\bm{\lambda}}^{c}]$, $[\ubar{\textbf{M}}]$, $[\ubar{\textbf{I}}^{uc}]$ and $[\ubar{\textbf{I}}^{c}]$. 
\State $k=0$; $K=0$
\For{$i=1:N$}
\State $K_i=\text{len}(\textbf{Z}_{ii})$
\State $k+=k_i$; $K+=K_i$
\State $\textbf{R}_{ii}=\text{real}(\textbf{Z}_{ii})$; $\textbf{X}_{ii}=\text{imag}(\textbf{Z}_{ii})$
\State $[{\textbf{I}}_{i},{\bm{\lambda}}_i]=\text{eig}(\textbf{X}_{ii},\textbf{R}_{ii})$
\State $\ubar{\textbf{I}}_{i}={\textbf{I}}_{i}(:,1:k_i)$
\EndFor
\State $[\textbf{I}^{uc}]=\text{zeros}([K,k])$; $[\textbf{Z}^c]=\text{zeros}([K,K])$
\State $k_i^{start}=1$; $k_i^{end}=k_1$
\State $K_i^{start}=1$; $K_i^{end}=K_1$
\For{$i=1:N$}
\State $[\textbf{I}^{uc}](K_i^{start}:K_i^{end},k_i^{start}:k_i^{end})=[\ubar{\textbf{I}}_{i}]$
\State $K_j^{start}=1$; $K_j^{end}=K_1$
\For{$j=1:N$}
\If{$i<=j$}
\State $[\textbf{Z}^c](K_i^{start}:K_i^{end},K_j^{start}:K_j^{end})=[\textbf{Z}_{ij}]$
\Else
\State $[\textbf{Z}^c](K_i^{start}:K_i^{end},K_j^{start}:K_j^{end})=[\textbf{Z}_{ji}]^T$
\EndIf
\State $K_j^{start}+=K_j$; $K_j^{end}+=K_{(j+1)}$
\EndFor
\State $k_i^{start}+=k_i$; $k_i^{end}+=k_{(i+1)}$
\State $K_i^{start}+=K_i$; $K_i^{end}+=K_{(i+1)}$
\EndFor

\State $\ubar{\textbf{P}}_{Zc}=[\ubar{\textbf{I}}^{uc}]^T[\textbf{Z}^c][\ubar{\textbf{I}}^{uc}]$ 
\State $\ubar{\textbf{P}}_{Rc}=\text{real}(\ubar{\textbf{P}}_{Zc})$; $\ubar{\textbf{P}}_{Xc}=\text{imag}(\ubar{\textbf{P}}_{Zc})$
\State $[\ubar{\textbf{M}},\ubar{\bm{\lambda}}^{c}]=\text{eig}(\ubar{\textbf{P}}_{Xc},\ubar{\textbf{P}}_{Rc})$
\State $\ubar{\textbf{I}}^{c}=[\ubar{\textbf{I}}^{uc}][\ubar{\textbf{M}}]$
\end{algorithmic}
\label{alg2}
\end{algorithm} 
\section{Numerical Results}
In order to understand the coupled characteristic mode and their coupling behaviors, we have considered the most elementary antenna: the dipole. The dipoles are chosen because the modal behaviors of dipoles are well-known and extensively studied in the literature. At first, we will discuss about coupled characteristic modes of a two-element array. Then five five-element dipole array has been discussed.
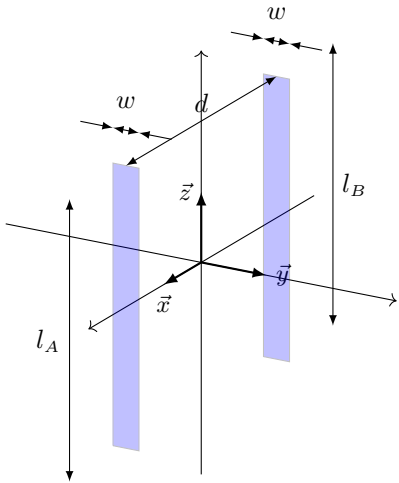
\begin{figure}
\centering
\tdplotsetmaincoords{70}{120}
\begin{tikzpicture}[tdplot_main_coords,scale=1]
\tikzstyle{every node}=[font=\small]
\draw[thick,-latex] (0,0,0) -- (1,0,0) node[below]{$\vec{x}$};
\draw[thick,-latex] (0,0,0) -- (0,1,0) node[right]{$\vec{y}$};
\draw[thick,-latex] (0,0,0) -- (0,0,1) node[left]{$\vec{z}$};
\draw[->] (-3,0,0) -- (3,0,0);
\draw[->] (0,-3,0) -- (0,3,0);
\draw[->] (0,0,-3) -- (0,0,3);
\filldraw[fill=blue, nearly transparent] (2,-0.2,-2) -- (2,0.2,-2) --  (2,0.2,2) -- (2,-0.2,2) -- (2,-0.2,-2);
\filldraw[fill=blue, nearly transparent] (-2,-0.2,-2) -- (-2,0.2,-2) --  (-2,0.2,2) -- (-2,-0.2,2) -- (-2,-0.2,-2);
\draw [dimen] (3.5,0,-2) -- (3.5,0,2) node[fill,opacity=0,text opacity=1,left] {$l_A$};
\draw [dimen] (2,0,2.0) -- (-2,0,2.0) node[fill,opacity=0,text opacity=1,above] {$d$};

\draw [dimen, ] (2,-0.2,2.5) -- (2,0.2,2.5) node [above=5pt] {$w$};
        \draw [dimen,<-] (2,-0.2,2.5) -- ++(0,-0.5,0);
        \draw [dimen,<-] (2,0.2,2.5) -- ++(0,0.5,0);

\draw [dimen] (-3.5,0,-2) -- (-3.5,0,2) node[fill,opacity=0,text opacity=1,right] {$l_B$};

\draw [dimen, ] (-2,-0.2,2.5) -- (-2,0.2,2.5) node [above=5pt] {$w$};
        \draw [dimen,<-] (-2,-0.2,2.5) -- ++(0,-0.5,0);
        \draw [dimen,<-] (-2,0.2,2.5) -- ++(0,0.5,0);
\end{tikzpicture}
\caption{Two element dipole array with $d=0.3$, $w=1/200$, $l_A=0.5$ and $l_B=0.3/0.5/0.7$ times wavelength.  }
\label{fig:two_ele_array}
\end{figure}
\subsection{Two Elements Dipole Array}
\begin{figure}
\includegraphics[width=8cm]{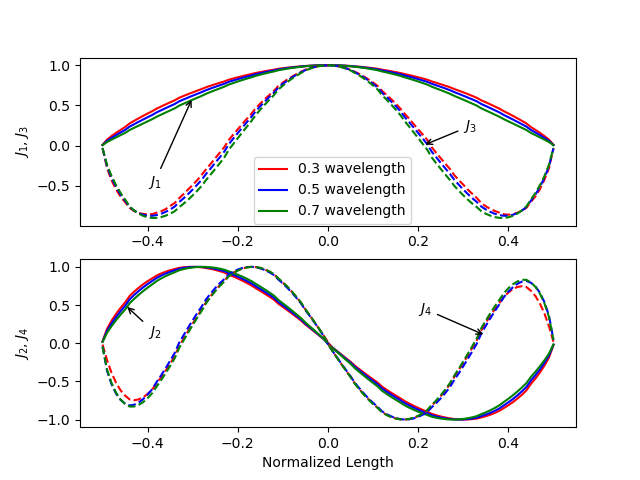}
\caption{First four isolated modes of three dipoles of length 0.3, 0.5 and 0.7 times wavelengths}
\label{fig:dipole_modes}
\end{figure}

In this example, two dipoles of different length combinations are considered. Lengths of dipoles are considered to be 0.3, 0.5 and 0.7 times the wavelength at the design frequency. At first isolated modes of the dipoles are calculated using EMPy: a python module developed at Circuit-EM Co-Design Lab. The modal currents $J_1$, $J_2$, $J_3$ and $J_4$ are plotted in Fig. \ref{fig:dipole_modes}, and the eigenvalues $\lambda_1$, $\lambda_2$, $\lambda_3$ and $\lambda_4$ are tabulated in Table-\ref{tab:iso_mode}. The modal currents $J_k$ with $k\in \{1,2,3,4\}$ in Fig. \ref{fig:dipole_modes} remain almost unaltered for different lengths of the dipoles. However, eigenvalues in Table-\ref{tab:iso_mode} are greatly affected by the length of the dipoles. It is interesting to note that the $4^{th}$ mode of dipole $l=0.3$ has a highly positive eigenvalue while as the length increases to $l=0.5$ the eigenvalue becomes highly negative. The result indicates discontinuity of higher order eigenvalue at low frequency. 

To demonstrate the coupled characteristic mode, we have consider three combination of dipole coupling considering the dipole are separated by 0.3 times of wavelength as shown in Fig. \ref{fig:two_ele_array}. The combination are Combo1: $(A=0.5, B=0.3)$, Combo2: $(A=0.5, B=0.5)$ and Combo3: $(A=0.5, B=0.7)$. The modal coupling analysis between first four mode of A i.e., $[\ubar{\textbf{I}}_A]=[I_1,I_2,I_3,I_4]_A$ and the first four modes of B i.e., $[\ubar{\textbf{I}}_B]=[I_1,I_2,I_3,I_4]_B$ has been carried out using Algorithm-\ref{alg1}. Coupled eigenvalues are predicted using Algorithm-\ref{alg1} is shown in Table-\ref{tab:coupled_eig_val}. The modal coupling matrices for different combinations of coupled dipoles are shown in Fig. \ref{fig:Coupling_matrix}. The nomenclature of coupled characteristic modes is based on the maximum contribution from the isolated characteristic mode. However, for identical dipoles $l_A=l_B=0.5$, the coupled modes always appear in an even-odd mode pair, with equal contribution from modes of A and B. In this case, lower eigenvalues have been associated with A and higher eigenvalues with B.
   
\begin{table}
\caption{Eigenvalue of first four isolated modes of three dipoles of length 0.3, 0.5 and 0.7 times wavelengths}
\centering
\begin{tabular}{|c|c|c|c|c|}
\hline\hline
Dipole & $\lambda_1$ & $\lambda_2$ & $\lambda_3$ & $\lambda_4$\\\hline
0.3 & $-12.49$ & $-1450.66$ & $-231091.71$ & $2.11\times10^9$\\\hline
0.5 & 0.67 & $-118.35$ & $-7457.49$ & $-768036.23$\\\hline
0.7 & 2.68 & $-18.22$ & $-753.78$ &
 $-40019.04$\\\hline
\end{tabular}
\label{tab:iso_mode}
\end{table}
The coupled eigenvalues are formed by perturbation of isolated eigenvalues due to mutual coupling. For example in Combo-1 $\lambda_{1A}=0.67$ perturbed to $\lambda^c_{1A}=0.71$, and  $\lambda_{1B}=-12.49$ to $\lambda^c_{1B}=-14.34$. It is to be noted that the amount of perturbation is greatly depends upon the coupling or perturbing structure. Electrically large coupling structures perturb the eigenvalue more significantly compared to electrically small structures. For example $\delta \lambda_{1A}=|\lambda^c_{1A}-\lambda_{1A}|=0.04$ and $\delta \lambda_{1A}=|\lambda^c_{1A}-\lambda_{1A}|=0.19$ for $l_B=0.3$ and $l_B=0.7$ respectively. However, maximum perturbation occurs when dipoles are identical, i.e.,  $\delta \lambda_{1A}=0.52$ and $\delta \lambda_{1B}=1.22$ for $l_A=l_B=0.5$. 
\begin{table}
\caption{Eigenvalues of Coupled Dipole Array}
\centering
\begin{tabular}{|p{0.75cm}|p{0.5cm}|p{1cm}|p{1cm}|p{1.3cm}|p{1.6cm}|}
\hline\hline
Combo & Asso. & $\lambda_1^c$ & $\lambda_2^c$ & $\lambda_3^c$ & $\lambda_4^c$\\\hline
\multirow{2}{1pt}{A=0.5,  B=0.3}&A& $0.71$ & $-115.43$ & $-7568.23$ & $-802075.13$ \\\cline{2-6}
&B& $-14.38$ & $-2142.58$ & $-380824.23$ & $123433553.46$ \\\hline

\multirow{2}{1pt}{A=0.5,  B=0.5}&A& $0.14$ & $-77.08$ & $-4707.65$ & $-478268.66$ \\\cline{2-6}
&B& $1.89$ & $-259.75$ & $-19951.11$ & $-2429974.08$ \\\hline

\multirow{2}{1pt}{A=0.5,  B=0.7}&A& $0.48$ & $-174.62$ & $-12688.53$ & $-1445198.57$ \\\cline{2-6}
&B& $3.91$ & $-17.47$ & $-743.56$ & $-41008.36$ \\\hline
\end{tabular}
\label{tab:coupled_eig_val}
\end{table}

The coupled characteristic modes can be fully described using modal coupling matrix $[\ubar{\textbf{M}}]$ and isolated modes $[\ubar{\textbf{I}}_A]$ and $[\ubar{\textbf{I}}_B]$. The modal coupling matrix $[\ubar{\textbf{M}}]$ should be read column-wise, where each column is associated with a coupled characteristic mode. For example consider the first column of $[\ubar{\textbf{M}}]$ in Fig. \ref{fig:Coupling_matrix}(a), which represents coupled mode $[I^c_{1A}]=\begin{bmatrix}
1000I_{1A}+0.1I_{3A}\\
90.64I_{1B}+0.15I_{3B}
\end{bmatrix}$. The coupled mode is formed due to the maximum contribution of $1^{st}$ isolated mode of element A, which is coupled to the $1^{st}$ mode of B. The first mode of A is also coupled to the third mode of B with negligible coupling. The third mode of B excites the third mode of A. Please note that the second and fourth modes do not contribute to the coupling between the first modes of A and B. As the second and fourth modes are totally dissimilar from the first mode. The second column of $[\ubar{\textbf{M}}]$ in Fig. \ref{fig:Coupling_matrix}(a) represents coupled mode $[I^c_{1B}]=\begin{bmatrix}
226.35I_{1A}-1.72I_{3A}\\
1000I_{1B}-0.08I_{3B}
\end{bmatrix}$, where the first mode of B provides maximum contribution. It is interesting to note that all the coupled modes appear in even-odd mode pairs when elements are identical as shown in Fig. \ref{fig:Coupling_matrix}(b). In the second mode coupling the fourth mode also contributes or vice versa.

After looking into all the mode coupling matrices we can conclude that one-to-one mode coupling is good enough to characterise modal coupling behaviour in dipole-like 1d structure. In the following subsection, we will characterise the coupled characteristic mode of the five-element identical array.       
\begin{figure}
\allowdisplaybreaks
\subfloat[]{
\begin{tikzpicture}[scale=0.8]
\draw (0,0) node[scale=0.7] {$
     \bordermatrix{ & A_1 & B_1 & A_2 & B_2 & A_3 & B_3 & A_4 & B_4 \cr
       A_1 & 1000	&226.35&	0&	0&	4.65&	2.77&	0&	0\cr
A_2 &0&	0&	1000&	224.84&	0&	0&	3.3&	0.86\cr
A_3 &0.1&	-1.72	&0	&0.01	&1000	&141.05&	0&	0\cr
A_4 &0&	0	&0	&-7.68	&0&	-0.02	&1000	&80.64\cr
B_1&90.64&	1000	&0.01&	0	&18.6&	3.95&	0&	0\cr
B_2&0&	0	&-121.38&	1000&	-0.01&	0&	15.63&	2.11\cr
B_3&0.15&	-0.08	&0	&0	&-89.47&	1000	&0.02&	0\cr
B_4&0	&0&	-0.02&	-0.34&	0&	0&	-62.39&	1000} \qquad
 $ };
\end{tikzpicture}}\\
\subfloat[]{
\begin{tikzpicture}[scale=0.8]
\draw (0,0) node[scale=0.7] {$
     \bordermatrix{ & A_1 & B_1 & A_2 & B_2 & A_3 & B_3 & A_4 & B_4 \cr
     A_1 &  1000&1000&0&0&-6.25&14.54&0&0 \cr
A_2 &0&0&1000&1000&0&0&-2.45&8.39 \cr
A_3 &-0.77&1.23&0&0&1000&1000&0&0 \cr
A_4 &0&0&0.15&-1.77&0&0&1000&1000 \cr
B_1 &-1000&1000&0&0&6.25&14.54&0&0 \cr
B_2 &0&0&-1000&1000&0&0&2.45&8.39 \cr
B_3 &0.77&1.23&0&0&-1000&1000&0&0 \cr
B_4 &0&0&-0.15&-1.77&0&0&-1000&1000 } \qquad
 $ };
\end{tikzpicture}}\\
\subfloat[]{
\begin{tikzpicture}[scale=0.8]
\draw (0,0) node[scale=0.7] {$
     \bordermatrix{ & A_1 & B_1 & B_2 & A_2 & B_3 & A_3 & B_4 & A_4 \cr
     A_1&1000&792.47&-0.03&0&30.19&10.97&0&0\cr
A_2&0&0&-154.81&1000&0.02&0&22.54&6.62\cr
A_3&-0.08&1.29&0&0&-174.37&1000&0&0\cr
A_4&0&0&0.14&-0.66&0&0&-135.42&1000\cr
B_1&-227.73&1000&-0.01&0.01&8.21&8.78&0&0\cr
B_2&-0.01&-0.01&1000&335.4&0&0&6.58&3.8\cr
B_3&2.49&4.9&0&-0.01&1000&249.89&0&0\cr
B_4&0&0&-0.13&-7.12&0&0&1000&182.25} \qquad
 $ };
\end{tikzpicture}
}
\caption{Modal coupling matrix of two element dipole array (a) Combo1: $(A=0.5, B=0.3)$, (b) Combo2: $(A=0.5, B=0.5)$ and (c) Combo3: $(A=0.5, B=0.7)$ }
\label{fig:Coupling_matrix}
\end{figure}
\begin{figure}
\centering
\tdplotsetmaincoords{70}{120}
\begin{tikzpicture}[tdplot_main_coords,scale=1]
\tikzstyle{every node}=[font=\small]
\draw[thick,-latex] (0,0,0) -- (1,0,0) node[below]{$\vec{x}$};
\draw[thick,-latex] (0,0,0) -- (0,1,0) node[right]{$\vec{y}$};
\draw[thick,-latex] (0,0,0) -- (0,0,1) node[left]{$\vec{z}$};
\draw[->] (-3,0,0) -- (3,0,0);
\draw[->] (0,-3,0) -- (0,3,0);
\draw[->] (0,0,-3) -- (0,0,3);
\filldraw[fill=blue, nearly transparent] (2,-0.2,-2) -- (2,0.2,-2) --  (2,0.2,2) -- (2,-0.2,2) -- (2,-0.2,-2);
\filldraw[fill=blue, nearly transparent] (-2,-0.2,-2) -- (-2,0.2,-2) --  (-2,0.2,2) -- (-2,-0.2,2) -- (-2,-0.2,-2);
\filldraw[fill=blue, nearly transparent] (0,-0.2,-2) -- (0,0.2,-2) --  (0,0.2,2) -- (0,-0.2,2) -- (0,-0.2,-2);
\filldraw[fill=blue, nearly transparent] (-1,-0.2,-2) -- (-1,0.2,-2) --  (-1,0.2,2) -- (-1,-0.2,2) -- (-1,-0.2,-2);
\filldraw[fill=blue, nearly transparent] (1,-0.2,-2) -- (1,0.2,-2) --  (1,0.2,2) -- (1,-0.2,2) -- (1,-0.2,-2);
\draw [dimen] (3.5,0,-2) -- (3.5,0,2) node[fill,opacity=0,text opacity=1,left] {$l$};
\draw [dimen] (-1,0,2.0) -- (-2,0,2.0) node[fill,opacity=0,text opacity=1,above] {$d$};
\draw [dimen, ] (2,-0.2,2.5) -- (2,0.2,2.5) node [above=5pt] {$w$};
        \draw [dimen,<-] (2,-0.2,2.5) -- ++(0,-0.5,0);
        \draw [dimen,<-] (2,0.2,2.5) -- ++(0,0.5,0);
%
%\draw [dimen] (-3.5,0,-2) -- (-3.5,0,2) node[fill,opacity=0,text opacity=1,right] {$l_B$};
%
%\draw [dimen, ] (-2,-0.2,2.5) -- (-2,0.2,2.5) node [above=5pt] {$w$};
%        \draw [dimen,<-] (-2,-0.2,2.5) -- ++(0,-0.5,0);
%        \draw [dimen,<-] (-2,0.2,2.5) -- ++(0,0.5,0);
\end{tikzpicture}
\caption{Five element dipole array with $d=0.4$, $w=1/200$ and $l=0.5$ times wavelength.  }
\label{fig:five_ele_array}
\end{figure}
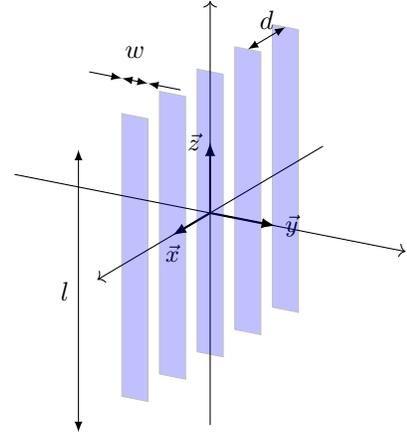
\subsection{Five Element Dipole Array  }
\begin{figure}
\includegraphics[width=9 cm]{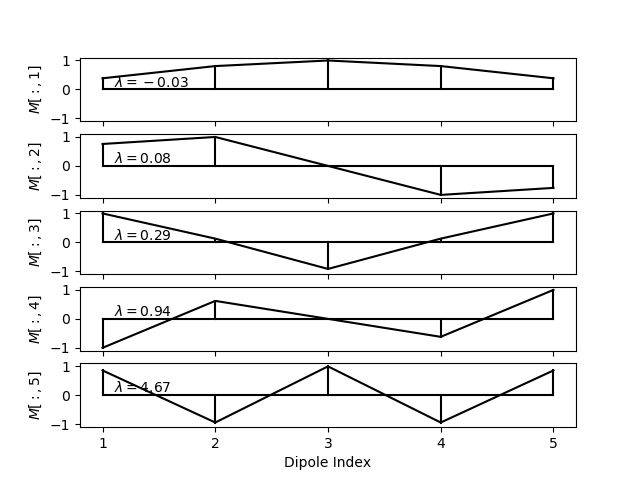}
\caption{Mode Coupling Pattern of the first mode of five element dipole array}
\label{fig:mode_coupling1}
\end{figure}

\begin{figure}
\includegraphics[width=9 cm]{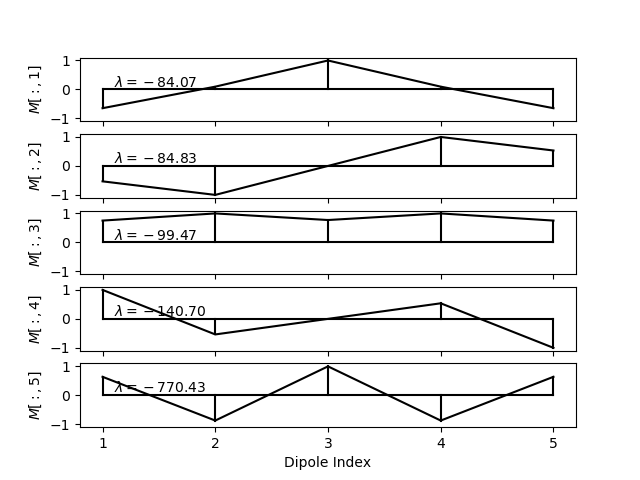}
\caption{Mode Coupling Pattern of the second mode of five element dipole array}
\label{fig:mode_coupling2}
\end{figure}

This example has been chosen to recreate the work of Inagaki and Garbacz in \cite{Inagaki1982}, where eigenfunctions and eigenfields of five element isotopic radiators array separated by $d=0.4$ times wavelength were determined. Here we have considered an array of five element identical dipoles of length $l=0.5$ times wavelength, and they are separated by $d=0.4$ times wavelength as shown in Fig. \ref{fig:five_ele_array}. The modal coupling matrix and coupled eigenvalues of the array have been determined considering only coupling between the first mode of each array element. The Algorithm-\ref{alg2} is deployed to determine the coupled modes. The mode coupling matrix is given as    
\begin{align}
&\textbf{M}_{1\rightleftharpoons 1}  =\nonumber\\&
     \bordermatrix{ & 1 & 2 & 3 &4&5\cr
       1 & 385.73	&757.92	&1000&	-1000&	859.61 \cr
2&807.35&	1000&	130.36	&620.33	&-938.14\cr
3&1000&	0	&-916.18&	0	&1000\cr
4&807.35&	-1000	&130.36&	-620.33	&-938.14\cr
5&385.73&	-757.92&	1000&	1000&	859.61}.\label{eq:M11}
\end{align}
The coupled eigenvalues and modal coupling pattern is shown in Fig. \ref{fig:mode_coupling1}. The mode coupling pattern presented in Fig. \ref{fig:mode_coupling1} is similar to \cite[Fig.2]{Inagaki1982}. Please note that for two two-element identical arrays for any fundamental mode, the modal coupling appears in even-odd pattern. However, the modal coupling patterns are different depending upon the choice of fundamental mode for more than two element arrays. For example,e mode coupling matrix of second mode coupling can be given as 
\begin{align}
&\textbf{M}_{2\rightleftharpoons 2}= \nonumber\\& 
     \bordermatrix{ & 1 & 2 & 3 &4&5\cr
       1 & -646.62&	-532.28&	752.92&	1000&	639.83 \cr
2&96.9	&-1000	&1000	&-537.45	&-870.54\cr
3&1000	&0	&772.66	&0	&1000\cr
4&96.9	&1000&	1000	&537.45&	-870.54\cr
5&-646.62	&532.28	&752.92	&-1000&	639.83}.\label{eq:M22}
\end{align}    
One can easily understand that columns of \eqref{eq:M22} are quite different from the columns of \eqref{eq:M11}. For better understanding, we can compare the mode coupling pattern shown in Fig. \ref{fig:mode_coupling1} and \ref{fig:mode_coupling2} for first and second mode coupling respectively. A details case study of one-to-one mode coupling of identical arrays have been provided in the supplementary file\footnote[2]{A supplementary document of this paper is available online at http://ieeexplore.ieee.org \label{note1}}. 
\section{Conclusions}
A general theory of coupled characteristic mode has been presented. The theory is based on the expansion of coupled eigen-subspace using isolated eigen-subspace. The eigen-subspace analysis is quite helpful in explaining current density, field pattern, and impedance matching of antenna and scatterer. The theory of two elements multi-mode coupling is established based on the assumption that uncoupled eigen-subspace can be mapped into coupled eigen-subspace via linear transform. The mapping matrix is named as modal coupling matrix. The mode coupling matrix is the measure of coupling among the isolated modes. The theory of two elements coupled mode has been extended to multi-element coupled characteristic mode. Two algorithms have been developed to calculate coupled eigenvalues, eigenvectors, and mode coupling matrix. The numerical examples conclude that maximum modal coupling occurs when elements are identical. Similar modes provide maximum modal coupling, while distinct modes are rarely coupled to each other.   For two-element identical elements, the coupled modes appear in an even-odd pair, while the same is not valid in the case of more than two elements.

\appendices

\renewcommand{\thefigure}{A\arabic{figure}}
\setcounter{figure}{0}
\renewcommand{\thetable}{A\arabic{table}} 
\setcounter{table}{0}
\renewcommand{\theequation}{A\arabic{equation}} 
\setcounter{equation}{0}

\section{Generalized Eigenvalue Problem}
Consider the generalized eigenvalue problem of characteristic mode 
\begin{equation}
[\textbf{X}][I_i]=\lambda_i[\textbf{R}][I_i] 
\end{equation} 
where $i=1,2,3,...,N$ and $N\times N$ is the dimension of $[\textbf{R}]$ and $[\textbf{X}]$. In this problem we have to find find characteristic mode $[I_i]$ and eigenvalue $\lambda_i$ of the $i^{th}$ mode. We can write the eigenvalue problem in matrix form as 
 \begin{equation}
[\textbf{X}][\textbf{I}]=[\textbf{R}][\textbf{I}] [\bm{\lambda}]
\end{equation} 
where $\textbf{I}=[I_1,I_2,I_3,...,I_N]$ and $\bm{\lambda}=diag\{\lambda_1,\lambda_2,...,\lambda_N\}$. To understand the properties of generalized eigenvalue problem, we need investigate the problem from the eigenvalue problem of $[\textbf{R}]$ and $[\textbf{X}]$. 

The normal eigenvalue problem of $[\textbf{R}]$ can be written as 
\begin{equation}
[\textbf{R}][\textbf{I}_R]=[\textbf{I}_R][\bm{\lambda}_R]\; or\;[\textbf{I}_R]^{T} [\textbf{R}][\textbf{I}_R]=[\textbf{I}^2_R][\bm{\lambda}_R].
\end{equation} 
Please note that $[\textbf{I}^2_R]=[\textbf{I}_R]^T[\textbf{I}_R]=diag\{||I_{R1}||^2,||I_{R2}||^2,...,||I_{RN}||^2\}$ is a diagonal matrix, not necessary to be identity. Now consider that the eigenvector space $[\textbf{I}]$ in the generalized eigenvalue problem is linear combination of eigenvector space $[\textbf{I}_R]$ through the following relationship
\begin{equation}
[\textbf{I}]=[\textbf{I}_R][\textbf{I}'_{X}].
\end{equation} 
Using this we can write 
\begin{equation}
[\textbf{X}][\textbf{I}_R][\textbf{I}'_{X}]=[\textbf{R}][\textbf{I}_R][\textbf{I}'_{X}] [\bm{\lambda}]
\end{equation}
After taking the inner product with respect to $[\textbf{I}_R]$ on both side of  
\begin{equation}
[\textbf{I}_R]^T[\textbf{X}][\textbf{I}_R][\textbf{I}'_{X}]=[\textbf{I}^2_R][\bm{\lambda}_R][\textbf{I}'_{X}] [\bm{\lambda}]
\end{equation}
which can be written as normal eigenvalue problem 
\begin{equation}
[\textbf{X}'][\textbf{I}'_{X}]=[\textbf{I}'_{X}] [\bm{\lambda}]\label{eq:eigx}
\end{equation}
where $[\textbf{X}']=[\bm{\lambda'}^{-1}_R][\textbf{I}_R]^T[\textbf{X}][\textbf{I}_R]$ and $[\bm{\lambda'}_R]=diag\{{\lambda_{R1}||I_{R1}||^2}, {\lambda_{R2}||I_{R2}||^2},...,{\lambda_{RN}||I_{RN}||^2}\}$. By solving \eqref{eq:eigx} we can get eigenvalues $[\bm{\lambda}]$, and eigenvectors $[\textbf{I}]=[\textbf{I}_R][\textbf{I}'_{X}]$ can be obtained by matrix multiplication. Please note that $[\textbf{X}']$ is not symmetric, i.e., $[\textbf{X}']^T\neq [\textbf{X}']$ or $[\textbf{I}_R]^T[\textbf{X}][\textbf{I}_R][\bm{\lambda'}^{-1}_R]\neq [\bm{\lambda'}^{-1}_R][\textbf{I}_R]^T[\textbf{X}][\textbf{I}_R]$. Therefore, the eigenvectors $[\textbf{I}'_{X}]$ are not orthogonal or $[\textbf{I}'_{X}]^T[\textbf{I}'_{X}]$ is not a diagonal matrix. However the eigenvalues $[\bm{\lambda}]$ are real if $[\textbf{R}]$ is positive semidefinite. 

Now we are going to prove that eigenvalues $[\bm{\lambda}]$ are real if $[\textbf{R}]$ is positive semidefinite. Let $i^{th}$ eigenvalue $\lambda_i$ is complex and corresponding  eigenvector $[I'_{Xi}]$ is also complex. Therefore, \eqref{eq:eigx} can be written as  
\begin{equation}
[\textbf{X}'][{I}'_{Xi}]=\lambda_i[{I}'_{Xi}]\label{eq:eigx1}
\end{equation} 
Taking complex conjugate on both side of \eqref{eq:eigx1} we have
\begin{equation}
[\bar{\textbf{X}'}][\bar{{I}'}_{Xi}]=\bar{\lambda}_i[\bar{{I}'}_{Xi}] \hspace{1em} \text{or} \hspace*{1em} [{\textbf{X}'}][\bar{{I}'}_{Xi}]=\bar{\lambda}_i[\bar{{I}'}_{Xi}] \label{eq:eigx2}
\end{equation} 
$[{\textbf{X}'}]=[\bm{\lambda'}^{-1}_R][\textbf{I}_R]^T[\textbf{X}][\textbf{I}_R]$ is real because all the factored matrices are real.  Now taking inner product with vector $[\bm{\lambda'}_R][{I}'_{Xi}]$ on both side of \eqref{eq:eigx1},we obtain 
 \begin{align}
 \lambda_i[\bar{{I}'}_{Xi}]^T[\bm{\lambda'}_R][{I}'_{Xi}]=&[\bar{{I}'}_{Xi}]^T[\bm{\lambda'}_R][\textbf{X}'][{I}'_{Xi}]\nonumber\\
 =&([\textbf{X}']^T[\bm{\lambda'}_R][\bar{{I}'}_{Xi}])^T[{I}'_{Xi}]\nonumber\\
 =& ([\textbf{I}_R]^T[\textbf{X}][\textbf{I}_R][\bm{\lambda'}^{-1}_R][\bm{\lambda'}_R][\bar{{I}'}_{Xi}])^T[{I}'_{Xi}]\nonumber\\
 =& ([\bm{\lambda'}_R][\bm{\lambda'}^{-1}_R][\textbf{I}_R]^T[\textbf{X}][\textbf{I}_R][\bar{{I}'}_{Xi}])^T[{I}'_{Xi}]\nonumber\\
 =& ([\bm{\lambda'}_R][\textbf{X}']][\bar{{I}'}_{Xi}])^T[{I}'_{Xi}]\nonumber\\
 =& ([\bm{\lambda'}_R]\bar{\lambda}_i[\bar{{I}'}_{Xi}])^T[{I}'_{Xi}]\nonumber\\
 =& \bar{\lambda}_i[\bar{{I}'}_{Xi}]^T[\bm{\lambda'}_R][{I}'_{Xi}]
 \end{align}
 Thus $(\lambda_i-\bar{\lambda}_i)[{I}'_{Xi}]^H[\bm{\lambda'}_R][{I}'_{Xi}]=0$. One of the result of this equation is $\bar{\lambda}_i=\lambda_i$ or $\lambda_i$ is real if $\phi_i=[{I}'_{Xi}]^H[\bm{\lambda'}_R][{I}'_{Xi}]\neq 0$. If $[\textbf{R}]$ is positive definite, then  $\phi_i=[{I}'_{Xi}]^H[\bm{\lambda'}_R][{I}'_{Xi}]=\sum_{n=1}^N {\lambda_{Rn}||I_{Rn}||^2}|{I}_{Xi}(n)|^2>0$, which ensure the eigenvalues are real. 
 
 As the eigenvalues $[\bm{\lambda}]$ of \eqref{eq:eigx} are real for positive definite $[\textbf{R}]$ matrix, the eigenvectors $[\textbf{I}'_{X}]$ are real. Therefore, eigenvectors $[\textbf{I}]=[\textbf{I}_R][\textbf{I}'_{X}]$  are also real as $[\textbf{I}_R]$ is real.   
 
 Now we are going to prove that $[{I}'_{Xj}]^H[\bm{\lambda'}_R][{I}'_{Xi}]=0$ for $i\neq j$. Let us write write the $j^{th}$ eigenvalue problem of \eqref{eq:eigx}
 \begin{equation}
[\textbf{X}'][{I}'_{Xj}]=\lambda_j[{I}'_{Xj}]\label{eq:eigxj}
\end{equation} 
 Taking inner product with $[\bm{\lambda'}_R][{I}'_{Xj}]$ on both side of \eqref{eq:eigx1}, we have 
 \begin{align}
 \lambda_i[{{I}'}_{Xj}]^H[\bm{\lambda'}_R][{I}'_{Xi}]=&[{{I}'}_{Xj}]^H[\bm{\lambda'}_R][\textbf{X}'][{I}'_{Xi}]\nonumber\\
 =&([\textbf{X}']^H[\bm{\lambda'}_R][{{I}'}_{Xj}])^H[{I}'_{Xi}]\nonumber\\
 =& ([\textbf{I}_R]^T[\textbf{X}][\textbf{I}_R][\bm{\lambda'}^{-1}_R][\bm{\lambda'}_R][{{I}'}_{Xj}])^H[{I}'_{Xi}]\nonumber\\
 =& ([\bm{\lambda'}_R][\bm{\lambda'}^{-1}_R][\textbf{I}_R]^T[\textbf{X}][\textbf{I}_R][{{I}'}_{Xj}])^H[{I}'_{Xi}]\nonumber\\
 =& ([\bm{\lambda'}_R][\textbf{X}']][\bar{{I}'}_{Xj}])^H[{I}'_{Xi}]\nonumber\\
 =& ([\bm{\lambda'}_R]{\lambda}_j[{{I}'}_{Xj}])^H[{I}'_{Xi}]\nonumber\\
 =& {\lambda}_j[{{I}'}_{Xj}]^H[\bm{\lambda'}_R][{I}'_{Xi}]
\end{align}  
Thus
\begin{align}
(\lambda_i-\lambda_j)[{{I}'}_{Xj}]^H[\bm{\lambda'}_R][{I}'_{Xi}]=0.
\end{align}
If $\lambda_i\neq\lambda_j$, then $[{{I}'}_{Xj}]^H[\bm{\lambda'}_R][{I}'_{Xi}]=0$ or eigenvectors $[\textbf{I}'_{X}]$ are orthogonal with respect to $[\bm{\lambda'}_R]$, i.e., 
\begin{align}
[\textbf{I}'_{X}]^T[\bm{\lambda'}_R][\textbf{I}'_{X}]=[\bm{\Phi}]
\end{align} 
where $[\bm{\Phi}]=diag\{\phi_1,\phi_2,\hdots,\phi_N\}$. Using this property we can prove the $[\textbf{R}]$ and $[\textbf{X}]$ orthogonality of generalized eigenvectors $[\textbf{I}]$ as
\begin{align}
[\textbf{I}]^T[\textbf{R}][\textbf{I}]=& [\textbf{I}'_{X}]^T [\textbf{I}_R]^T [\textbf{R}][\textbf{I}_R][\textbf{I}'_{X}]\nonumber\\
=& [\textbf{I}'_{X}]^T [\bm{\lambda'}_R][\textbf{I}'_{X}]=[\bm{\Phi}]
\end{align}
and
\begin{align}
[\textbf{I}]^T[\textbf{X}][\textbf{I}]=[\textbf{I}]^T[\textbf{R}][\textbf{I}][\bm{\lambda}]=[\bm{\Phi}][\bm{\lambda}].
\end{align}
Thus properties of generalized eigenvalue problem are explained. 


%\bibliographystyle{IEEEtran}
\bibliographystyle{IEEEtran}
\bibliography{Ref2}

\begin{thebibliography}{10}
\providecommand{\url}[1]{#1}
\csname url@samestyle\endcsname
\providecommand{\newblock}{\relax}
\providecommand{\bibinfo}[2]{#2}
\providecommand{\BIBentrySTDinterwordspacing}{\spaceskip=0pt\relax}
\providecommand{\BIBentryALTinterwordstretchfactor}{4}
\providecommand{\BIBentryALTinterwordspacing}{\spaceskip=\fontdimen2\font plus
\BIBentryALTinterwordstretchfactor\fontdimen3\font minus
  \fontdimen4\font\relax}
\providecommand{\BIBforeignlanguage}[2]{{%
\expandafter\ifx\csname l@#1\endcsname\relax
\typeout{** WARNING: IEEEtran.bst: No hyphenation pattern has been}%
\typeout{** loaded for the language `#1'. Using the pattern for}%
\typeout{** the default language instead.}%
\else
\language=\csname l@#1\endcsname
\fi
#2}}
\providecommand{\BIBdecl}{\relax}
\BIBdecl

\bibitem{kouyoumjian1965asymptotic}
R.~G. Kouyoumjian, ``Asymptotic high-frequency methods,'' \emph{Proceedings of
  the IEEE}, vol.~53, no.~8, pp. 864--876, 1965.

\bibitem{keller1962geometrical}
J.~B. Keller, ``Geometrical theory of diffraction,'' \emph{Josa}, vol.~52,
  no.~2, pp. 116--130, 1962.

\bibitem{harrington1961time}
R.~F. Harrington, \emph{Time-Harmonic Electromagnetic Fields}.\hskip 1em plus
  0.5em minus 0.4em\relax New York: McGraw Hill, 1968.

\bibitem{felsen1994radiation}
L.~B. Felsen and N.~Marcuvitz, \emph{Radiation and scattering of waves}.\hskip
  1em plus 0.5em minus 0.4em\relax John Wiley \& Sons, 1994, vol.~31.

\bibitem{harrington1993field}
R.~F. Harrington, \emph{Field Computation by Moment Methods}.\hskip 1em plus
  0.5em minus 0.4em\relax New York: Macmillan Comp., 1968.

\bibitem{rao1982electromagnetic}
S.~Rao, D.~Wilton, and A.~Glisson, ``Electromagnetic scattering by surfaces of
  arbitrary shape,'' \emph{IEEE Trans. Antennas Propagat.}, vol.~30, no.~3, pp.
  409--418, 1982.

\bibitem{garbacz1965modal}
R.~Garbacz, ``Modal expansions for resonance scattering phenomena,''
  \emph{Proceedings of the IEEE}, vol.~53, no.~8, pp. 856--864, 1965.

\bibitem{garbacz1971generalized}
R.~Garbacz and R.~Turpin, ``A generalized expansion for radiated and scattered
  fields,'' \emph{IEEE Trans. Antennas Propag.}, vol.~19, no.~3, pp. 348--358,
  1971.

\bibitem{harrington1971theory}
R.~Harrington and J.~Mautz, ``Theory of characteristic modes for conducting
  bodies,'' \emph{IEEE Trans. Antennas Propag.}, vol.~19, no.~5, pp. 622--628,
  Sep 1971.

\bibitem{harrington1971computation}
------, ``Computation of characteristic modes for conducting bodies,''
  \emph{IEEE Trans. Antennas Propag.}, vol.~19, no.~5, pp. 629--639, 1971.

\bibitem{harrington1972characteristic}
R.~Harrington, J.~Mautz, and Y.~Chang, ``Characteristic modes for dielectric
  and magnetic bodies,'' \emph{IEEE Trans. Antennas Propagat.}, vol.~20, no.~2,
  pp. 194--198, 1972.

\bibitem{MATLAB}
MATLAB, \emph{ver. R2021a}.\hskip 1em plus 0.5em minus 0.4em\relax Natick,
  Massachusetts: The MathWorks Inc., 2021.

\bibitem{bekers2006eigencurrent}
D.~J. Bekers, S.~J. van Eijndhoven, A.~A. van~de Ven, P.-P. Borsboom, and A.~G.
  Tijhuis, ``Eigencurrent analysis of resonant behavior in finite antenna
  arrays,'' \emph{IEEE Trans. on Microw. Theory Techn.}, vol.~54, no.~6, pp.
  2821--2829, 2006.

\bibitem{newman1979small}
E.~Newman, ``Small antenna location synthesis using characteristic modes,''
  \emph{IEEE Trans. Antennas Propagat.}, vol.~27, no.~4, pp. 530--531, 1979.

\bibitem{bekers2009eigencurrent}
D.~J. Bekers, S.~J. van Eijndhoven, and A.~G. Tijhuis, ``An eigencurrent
  approach for the analysis of finite antenna arrays,'' \emph{IEEE Trans.
  Antennas Propag.}, vol.~57, no.~12, pp. 3772--3782, 2009.

\bibitem{cabedo2007theory}
M.~Cabedo-Fabres, E.~Antonino-Daviu, A.~Valero-Nogueira, and M.~F. Bataller,
  ``The theory of characteristic modes revisited: A contribution to the design
  of antennas for modern applications,'' \emph{IEEE Antennas and Propag. Mag.},
  vol.~49, no.~5, pp. 52--68, Oct 2007.

\bibitem{manteuffel2016compact}
D.~Manteuffel and R.~Martens, ``Compact multimode multielement antenna for
  indoor \text{UWB} massive \text{MIMO},'' \emph{IEEE Trans. Antennas Propag.},
  vol.~64, no.~7, pp. 2689--2697, 2016.

\bibitem{capek2012method}
M.~Capek, P.~Hazdra, and J.~Eichler, ``A method for the evaluation of radiation
  \text{Q} based on modal approach,'' \emph{IEEE Trans. Antennas Propag.},
  vol.~60, no.~10, pp. 4556--4567, Oct 2012.

\bibitem{wu2016reduction}
Q.~Wu, W.~Su, Z.~Li, and D.~Su, ``Reduction in out-of-band antenna coupling
  using characteristic mode analysis,'' \emph{IEEE Trans. Antennas Propagat.},
  vol.~64, no.~7, pp. 2732--2742, 2016.

\bibitem{rothenhausler2017characteristic}
M.~Rothenh{\"a}usler and F.~Gronwald, ``Characteristic mode analysis of
  \text{HIRF-and DCI}-excitations of an aircraft structure,'' in
  \emph{Electromagnetic Compatibility-EMC EUROPE, 2017 International Symposium
  on}.\hskip 1em plus 0.5em minus 0.4em\relax IEEE, 2017, pp. 1--6.

\bibitem{sandip_temc1}
S.~{Ghosal}, A.~{De}, A.~P. {Duffy}, and A.~{Chakrabarty}, ``Selection of
  dominant characteristic modes,'' \emph{IEEE Trans Electromagn Compat},
  vol.~62, no.~2, pp. 451--460, 2020.

\bibitem{gronwaldcharacteristic}
F.~Gronwald, ``Characteristic mode analysis as a pattern recognition technique
  for electromagnetic compatibility,'' in \emph{2nd URSI AT-RASC, Gran
  Canaria}.\hskip 1em plus 0.5em minus 0.4em\relax URSI, 2018.

\bibitem{sghosalthesis}
S.~Ghosal, ``Characteristic mode analysis for coupled systems,'' Ph.D.
  dissertation, Indian Institute of Technology Kharagpur, 2020.

\bibitem{raines2011systematic}
B.~D. Raines, ``Systematic design of multiple antenna systems using
  characteristic modes,'' Ph.D. dissertation, The Ohio State University, 2011.

\bibitem{ethier2012sub}
J.~Ethier and D.~McNamara, ``Sub-structure characteristic mode concept for
  antenna shape synthesis,'' \emph{Electron. Lett.}, vol.~48, no.~9, pp.
  471--472, 2012.

\bibitem{lou2018analysis}
S.~Lou, B.~Duan, W.~Wang, C.~Ge, and S.~Qian, ``Analysis of finite antenna
  arrays using the characteristic modes of isolated radiating elements,''
  \emph{IEEE Trans. Antennas Propagat.}, vol.~67, no.~3, pp. 1582--1589, 2018.

\bibitem{cheng2019novel}
G.~Cheng and C.-F. Wang, ``A novel periodic characteristic mode analysis method
  for large-scale finite arrays,'' \emph{IEEE Trans. Antennas Propagat.},
  vol.~67, no.~12, pp. 7637--7642, 2019.

\bibitem{kaffash2020fast}
S.~Kaffash, R.~Faraji-Dana, M.~Shahabadi, and S.~Safavi-Naeini, ``A fast
  computational method for characteristic modes and eigenvalues of array
  antennas,'' \emph{IEEE Trans. Antennas Propagat.}, vol.~68, no.~12, pp.
  7879--7892, 2020.

\bibitem{Oijala_TCCM_2019}
P.~Ylä-Oijala, A.~Lehtovuori, H.~Wallén, and V.~Viikari, ``Coupling of
  characteristic modes on pec and lossy dielectric structures,'' \emph{IEEE
  Trans. Antennas Propag.}, vol.~67, no.~4, pp. 2565--2573, 2019.

\bibitem{Lonsky_dipole_array_2018}
T.~Lonsky, P.~Hazdra, and J.~Kracek, ``Characteristic modes of dipole arrays,''
  \emph{IEEE Antennas Wireless Propag. Lett.}, vol.~17, no.~6, pp. 998--1001,
  2018.

\bibitem{Liang_Coupling_2018}
P.~Liang and Q.~Wu, ``Characteristic mode analysis of antenna mutual coupling
  in the near field,'' \emph{IEEE Trans. Antennas Propag.}, vol.~66, no.~7, pp.
  3757--3762, 2018.

\bibitem{tzanidis2012characteristic}
I.~Tzanidis, K.~Sertel, and J.~L. Volakis, ``Characteristic excitation taper
  for ultrawideband tightly coupled antenna arrays,'' \emph{IEEE Trans.
  Antennas Propagat.}, vol.~60, no.~4, pp. 1777--1784, 2012.

\bibitem{ghosal2020vertical}
S.~Ghosal, A.~De, R.~M. Shubair, and A.~Chakrabarty, ``Vertical dipole above
  the lossy dielectric half-space—a characteristic mode analysis,''
  \emph{IEEE Trans Electromagn Compat}, vol.~62, no.~6, pp. 2832--2841, 2020.

\bibitem{guo2020miniaturized}
Q.~Guo, J.~Su, Z.~Li, J.~Song, and Y.~Guan, ``Miniaturized-element
  frequency-selective rasorber design using characteristic modes analysis,''
  \emph{IEEE Trans. Antennas Propag.}, vol.~68, no.~9, pp. 6683--6694, 2020.

\bibitem{TCCM1}
S.~Ghosal, R.~Sinha, A.~De, A.~Chakrabarty, and H.~Son, ``Theory of coupled
  characteristic modes,'' \emph{IEEE Trans. Antennas Propag.}, vol.~68, no.~6,
  pp. 4677--4687, 2020.

\bibitem{TCCM2}
S.~Ghosal, R.~Sinha, A.~De, and A.~Chakrabarty, ``Characteristic mode analysis
  of mutual coupling,'' \emph{IEEE Trans. Antennas Propag.}, vol. Early Access,
  no.~xx, pp. 1--12, 2021.

\bibitem{ghosal2019vtc}
S.~Ghosal, A.~De, and A.~Chakrabarty, ``Eigenvalue based mutual coupling
  reduction,'' in \emph{2019 Proc. IEEE 89th Vehcl. Tech. Conference}.\hskip
  1em plus 0.5em minus 0.4em\relax IEEE, 04 2019.

\bibitem{ghosal2019characteristic}
S.~Ghosal, A.~De, A.~Chakrabarty, and R.~M. Shubair, ``A characteristic mode
  based decoupling approach,'' in \emph{2019 Proc. IEEE Int. Symp. Antennas
  Propag. USNC-URSI Radio Sci. Meeting}.\hskip 1em plus 0.5em minus 0.4em\relax
  IEEE, 2019, pp. 1--2.

\bibitem{ghosal2021incap}
S.~Ghosal, R.~Sinha, and A.~De, ``Further insights into coupled characteristic
  modes,'' in \emph{2021 Proc. IEEE Indian Conf. on Antennas and Propag.}\hskip
  1em plus 0.5em minus 0.4em\relax IEEE, 12 2021, pp. 1--4.

\bibitem{Inagaki1982}
N.~Inagaki and R.~Garbacz, ``Eigenfunctions of composite hermitian operators
  with application to discrete and continuous radiating systems,'' \emph{IEEE
  Trans. Antennas Propag.}, vol.~30, no.~4, pp. 571--575, 1982.

\end{thebibliography}

\end{document}

\title{Supplementary Document to General Theory of Coupled Characteristic Mode}
\author{ Rakesh~Sinha and Sandip~Ghosal}
\maketitle
\IEEEpeerreviewmaketitle
\section{Coupling Among Identical Elements}
To illustrate the first order interaction modeling of multiple antennas, half-wavelength PEC thin dipole having the width of 0..5 times the free space wavelength is considered. The  inter element spacing is assumed to be of 0.4 times the free-space wavelength. 
\subsection{ Two Dipoles}
Using the general theory of coupled characteristic modes (GTCCM), the coupling coefficients for two dipoles with one-to-one interaction modeling with identical modes of antenna 1 and 2, respectively, are given below for different choices of the fundamental mode.

% \begin{subequations}
% \allowdisplaybreaks
 \begin{align}
 \textbf{M}_{1\rightleftharpoons 1} & =
     \bordermatrix{ & 1 & 2 \cr
       1 & 1&1\cr
2& 1 & -1},~
\textbf{M}_{2\rightleftharpoons 2}=
     \bordermatrix{ & 1 & 2 \cr
       1 & 1&1\cr
2& 1 & -1},\\
\textbf{M}_{3\rightleftharpoons 3} & =
     \bordermatrix{ & 1 & 2 \cr
       1 & 1&1\cr
2& 1 & -1},~\text{and}~
\textbf{M}_{4\rightleftharpoons 4}  =
     \bordermatrix{ & 1 & 2 \cr
       1 & 1&1\cr
2& 1 & -1}
 \end{align}
% \end{subequations}
 
 \begin{figure*}[!h]
 \centering
 \subfloat[]{\includegraphics[width=9 cm, height=5 cm]{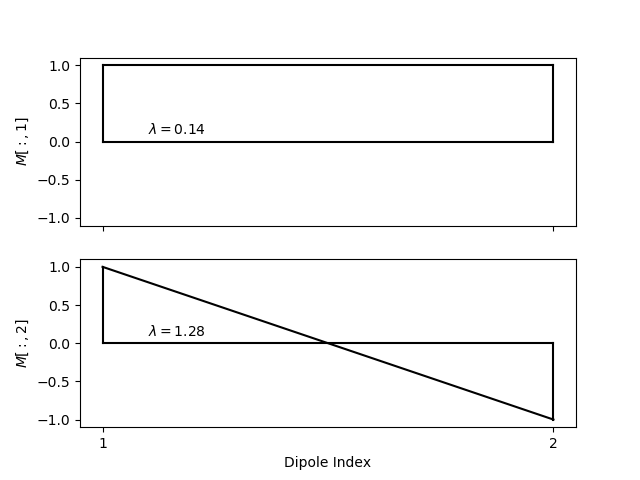}}
  \subfloat[]{\includegraphics[width=9 cm, height=5 cm]{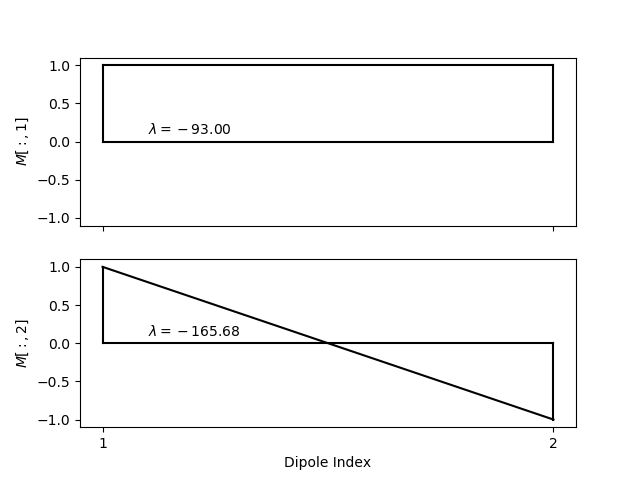}}\\
   \subfloat[]{\includegraphics[width=9 cm, height=5 cm]{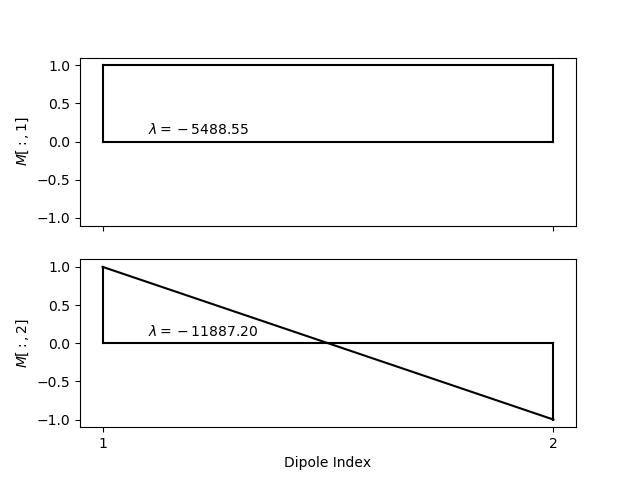}}
    \subfloat[]{\includegraphics[width=9 cm, height=5 cm]{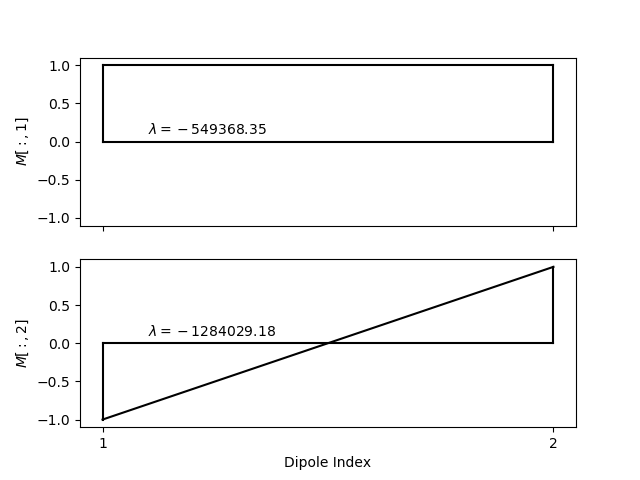}}
    
    \caption{Modal coefficients for one-to-one interaction with two dipoles (a) $1^{st}$ mode coupling (b)$2^{nd}$ mode coupling, (c) $3^{rd}$ mode coupling and (d) $4^{th}$ mode coupling.}
 \end{figure*}

\subsection{ Three Dipoles} 
 Using the GTCCM, the coupling coefficients for three dipoles with one fundamental mode coupling of antenna 1, 2, and 3, respectively, are given below for different choices of the fundamental mode.
\begin{subequations}
\allowdisplaybreaks
 \begin{align}
 \textbf{M}_{1\rightleftharpoons 1} & =
     \bordermatrix{ & 1 & 2 & 3\cr
       1 & 626.36&	1000&	1000 \cr
2&1000	&0&	-999.18\cr
3&626.36	&-1000	&1000},\\
 \textbf{M}_{2\rightleftharpoons 2}&  =
     \bordermatrix{ & 1 & 2 & 3\cr
       1 &503.09&	1000&	-987.32 \cr
2&1000	&0	&1000\cr
3&503.09	&-1000&	-987.32}\\
 \textbf{M}_{3\rightleftharpoons 3} & =
     \bordermatrix{ & 1 & 2 & 3\cr
       1 &610.49	&-1000&	-818.8 \cr
2&1000	&0	&1000\cr
3&610.49	&1000	&-818.8},\\
\textbf{M}_{4\rightleftharpoons 4}&  =
     \bordermatrix{ & 1 & 2 & 3\cr
       1 &661.25&	1000&	-756.13 \cr
2&1000	&0	&1000\cr
3&661.25&	-1000&	-756.13}
 \end{align}
 \end{subequations}
  \begin{figure*}[!h]
 \centering
 \subfloat[]{\includegraphics[width=9 cm]{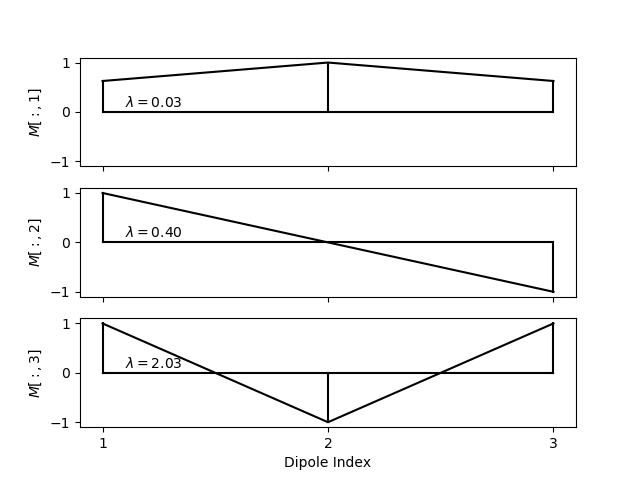}}
  \subfloat[]{\includegraphics[width=9 cm]{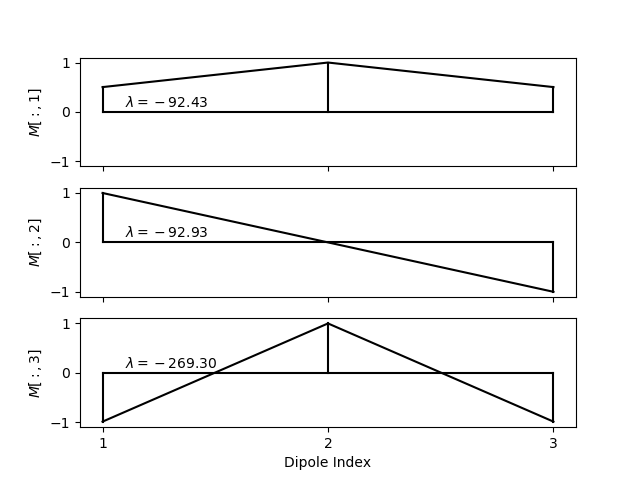}}\\
   \subfloat[]{\includegraphics[width=9 cm]{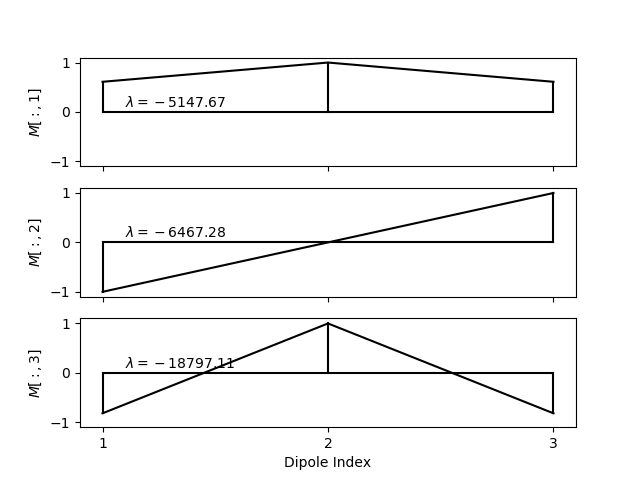}}
    \subfloat[]{\includegraphics[width=9 cm]{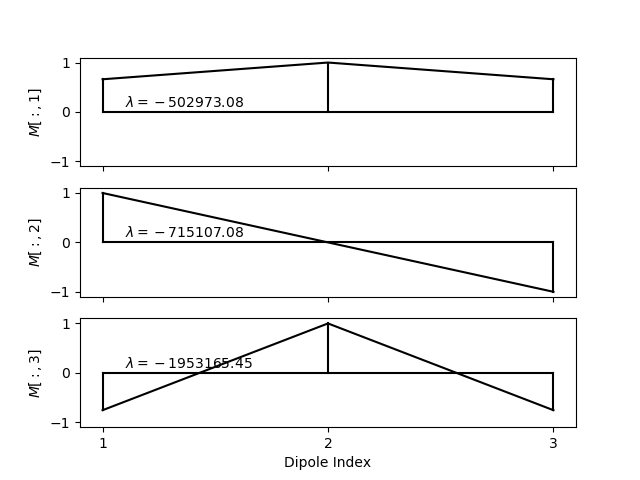}}
     \caption{Modal coefficients for one-to-one interaction with three dipoles (a) $1^{st}$ mode coupling (b)$2^{nd}$ mode coupling, (c) $3^{rd}$ mode coupling and (d) $4^{th}$ mode coupling.}
 \end{figure*}
 
 \subsection{Four Dipoles} 
 Using the GTCCM, the coupling coefficients for three dipoles with one fundamental mode coupling of antenna 1, 2, 3 and 4, respectively, are given below for different choices of the fundamental mode.
 \begin{small}
\begin{subequations}
\allowdisplaybreaks
 \begin{align}
 \textbf{M}_{1\rightleftharpoons 1} & =
     \bordermatrix{ & 1 & 2 & 3 &4\cr
       1 & 486.89	&-1000	&1000&	957.85 \cr
2&1000	&-708.45	&-419.55&	-1000\cr
3&1000&	708.45&	-419.55	&1000\cr
4&486.89&	1000&	1000&	-957.85},\\
 \textbf{M}_{2\rightleftharpoons 2} & =
     \bordermatrix{ & 1 & 2 & 3 &4\cr
       1 & -1000&	22.52&	1000&	810.3 \cr
2&-814.32	&1000&	-29.08&	-1000\cr
3&814.32&	1000	&-29.08	&1000\cr
4&1000	&22.52&	1000&	-810.3}, \\
 \textbf{M}_{3\rightleftharpoons 3} & =
     \bordermatrix{ & 1 & 2 & 3 &4\cr
       1 &424.04&	-1000&	1000&	-678.29 \cr
2&1000	&-678.39&	-424.28&	1000\cr
3&1000	&678.39&	-424.28&	-1000\cr
4&424.04&	1000&	1000&	678.29},\\
 \textbf{M}_{4\rightleftharpoons 4} & =
     \bordermatrix{ & 1 & 2 & 3 &4\cr
       1 & 565.01&	1000	&1000&	670.61 \cr
2&1000&	670.63	&-565.02	&-1000\cr
3&1000	&-670.63	&-565.02	&1000\cr
4&565.01&	-1000&	1000&	-670.61}. 
 \end{align}
 \end{subequations} 
 \end{small}
 \vspace{-1em}
 
 \begin{figure*}[!h]
 \centering
 \subfloat[]{\includegraphics[width=9 cm]{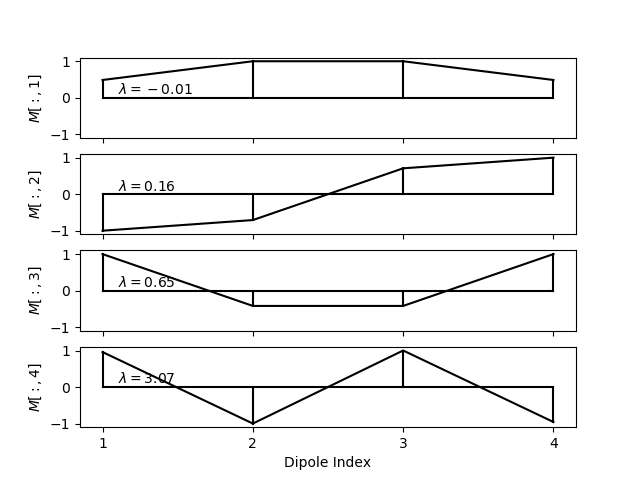}}
  \subfloat[]{\includegraphics[width=9 cm]{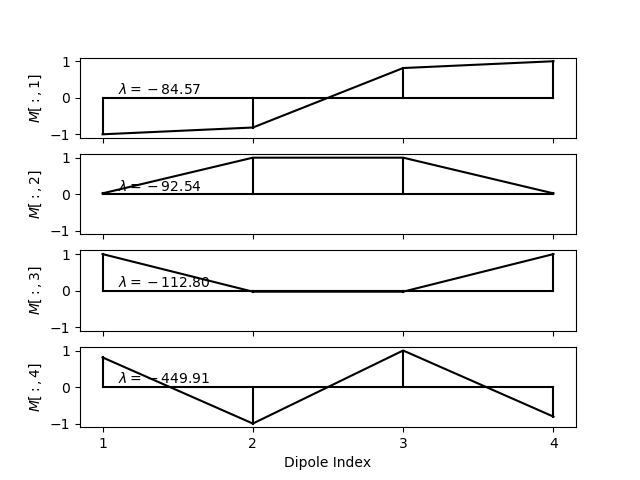}}\\
   \subfloat[]{\includegraphics[width=9 cm]{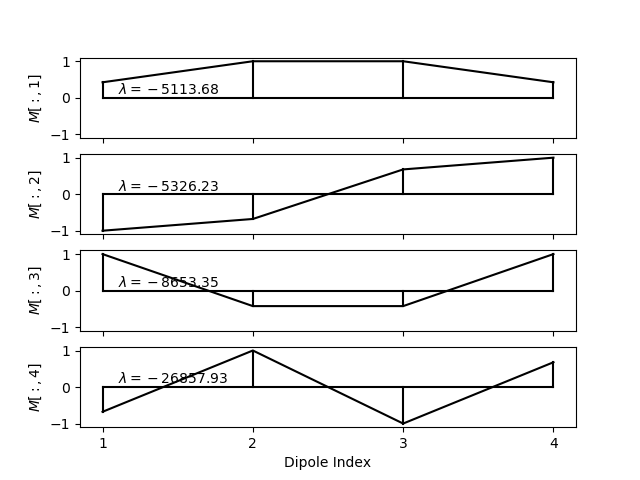}}
    \subfloat[]{\includegraphics[width=9 cm]{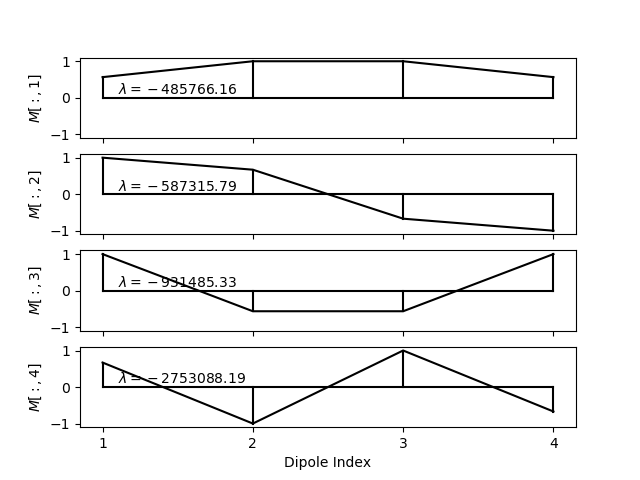}}
    \caption{Modal coefficients for one-to-one interaction with four dipoles (a) $1^{st}$ mode coupling (b) $2^{nd}$ mode coupling, (c) $3^{rd}$ mode coupling and (d) $4^{th}$ mode coupling.}
 \end{figure*}
 
 \subsection{ Five Dipoles} 
 Using the GTCCM, the coupling coefficients for three dipoles with one fundamental mode coupling of antenna 1, 2, 3, 4 and 5, respectively, are given below for different choices of the fundamental mode.
 \begin{small}
\begin{subequations}
\allowdisplaybreaks
 \begin{align}
 \textbf{M}_{1\rightleftharpoons 1} & =
     \bordermatrix{ & 1 & 2 & 3 &4&5\cr
       1 & 385.73	&757.92	&1000&	-1000&	859.61 \cr
2&807.35&	1000&	130.36	&620.33	&-938.14\cr
3&1000&	0	&-916.18&	0	&1000\cr
4&807.35&	-1000	&130.36&	-620.33	&-938.14\cr
5&385.73&	-757.92&	1000&	1000&	859.61}, \\
 \textbf{M}_{2\rightleftharpoons 2} & =
     \bordermatrix{ & 1 & 2 & 3 &4&5\cr
       1 & -646.62&	-532.28&	752.92&	1000&	639.83 \cr
2&96.9	&-1000	&1000	&-537.45	&-870.54\cr
3&1000	&0	&772.66	&0	&1000\cr
4&96.9	&1000&	1000	&537.45&	-870.54\cr
5&-646.62	&532.28	&752.92	&-1000&	639.83}, \\
\textbf{M}_{3\rightleftharpoons 3} & =
     \bordermatrix{ & 1 & 2 & 3 &4&5\cr
       1 & -872.36&	124.84&	1000&	-1000&	562.1 \cr
2&-1000&	670.49&	287.32&	872.52&	-850.47\cr
3&0	&1000&	-635.67&	0	&1000\cr
4&1000	&670.49&	287.32&	-872.52	&-850.47\cr
5&872.36&	124.84	&1000	&1000&	562.1}, \\
\textbf{M}_{4\rightleftharpoons 4} & =
     \bordermatrix{ & 1 & 2 & 3 &4&5\cr
       1 & 349.61	&1000	&1000&	925.23	&651.63 \cr
2&809.84&	925.21	&176.77	&-1000	&-898.73\cr
3&1000	&0	&-985.58	&0&	1000\cr
4&809.84	&-925.21	&176.77&	1000&	-898.73\cr
5&349.61	&-1000	&1000	&-925.23&	651.63}.
 \end{align}
 \end{subequations}
 \end{small}
 
 \begin{figure*}[!h]
 \centering
 \subfloat[]{\includegraphics[width=9 cm]{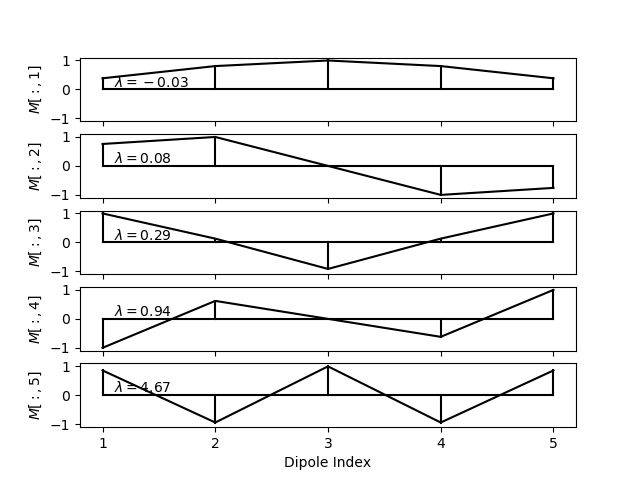}}
  \subfloat[]{\includegraphics[width=9 cm]{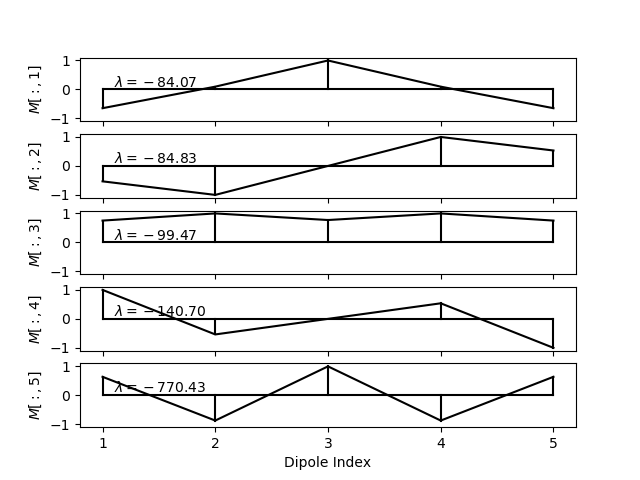}}\\
   \subfloat[]{\includegraphics[width=9 cm]{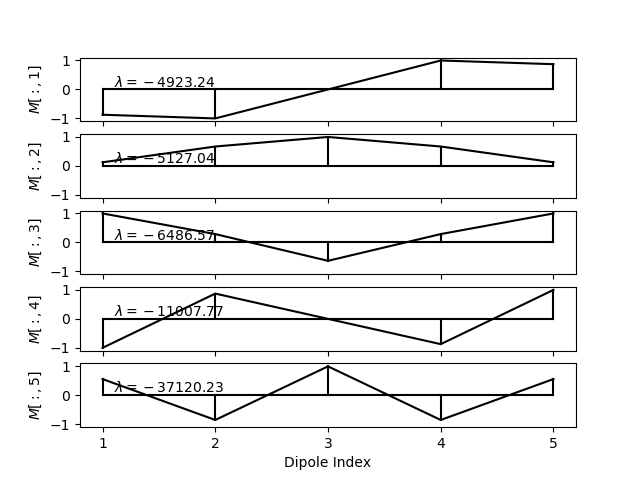}}
    \subfloat[]{\includegraphics[width=9 cm]{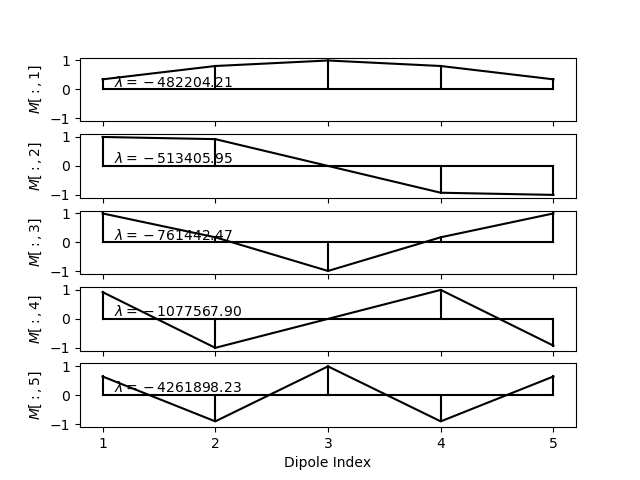}}
        \caption{Modal coefficients for one-to-one interaction with five dipoles (a) $1^{st}$ mode coupling (b) $2^{nd}$ mode coupling, (c) $3^{rd}$ mode coupling and (d) $4^{th}$ mode coupling.}
 \end{figure*}

%\bibliographystyle{IEEEtran}
%\bibliography{Ref2}